\documentclass[10pt,journal,compsoc]{IEEEtran}
\IEEEoverridecommandlockouts

\usepackage{cite}
\usepackage{amsmath,amssymb,amsfonts}
\usepackage{algorithmic}
\usepackage{textcomp}
\usepackage{xcolor}

\usepackage{booktabs} 

\usepackage[ruled]{algorithm2e} 

\SetAlFnt{\small}
\SetAlCapFnt{\small}
\SetAlCapNameFnt{\small}
\SetAlCapHSkip{0pt}
\IncMargin{-\parindent}

\usepackage{graphicx}

\usepackage{color}
\usepackage{url}
\usepackage{multirow}
\usepackage{rotating}

\usepackage{subfigure}

\newcommand{\commentout}[1]{}

\usepackage[normalem]{ulem}
\newcommand{\revrev}[2]{#2}

\newcommand{\revtwo}[2]{#2}

\newcommand{\revthree}[2]{#2}

\newcommand{\revExt}[2]{#2}

\newcommand{\revJournal}[2]{#2}

\newcommand{\revfour}[2]{#2}

\newcommand{\revDerek}[2]{#2}

\newcommand{\revfive}[2]{#2}

\usepackage{fancyhdr}
\usepackage{extramarks}

\begin{document}

\title{Optimized Dynamic Cache Instantiation and Accurate LRU Approximations under Time-varying Request Volume
  \thanks{A preliminary version of this paper appears as a 9-page paper at IFIP Networking 2020~\cite{CaEa20-networking}.}}

\IEEEaftertitletext{\vspace{-2\baselineskip}}

\author{\IEEEauthorblockN{Niklas Carlsson}
  \IEEEauthorblockA{\textit{Link\"oping University}, Sweden}
  \\
  \IEEEauthorblockN{Derek Eager}
  \IEEEauthorblockA{\textit{University of Saskatchewan}, Canada}
   }

\maketitle

\begin{abstract}

\revJournal{By caching content at geographically distributed servers,
content delivery applications can achieve scalability and reduce
wide-area network traffic.
However, each deployed cache has an associated cost.
When the request rate from the local region is sufficiently high
this cost will be justified, but as the request rate varies, for
example according to a daily cycle, there may be long periods when
the benefit of the cache does not justify the cost.
Cloud computing offers a solution to problems of this kind,
by supporting the dynamic allocation and release of
\revthree{resources according to need.}{resources.}
In this paper, we analyze the potential benefits from
dynamically instantiating caches using resources from cloud service providers.
We develop novel analytic caching models that accommodate time-varying
request rates, transient behavior as a cache fills following
instantiation, and selective cache insertion policies.
Using these models,
\revrev{and within}{within}
the context of a simple cost model,
\revrev{we are able}{we then develop bounds and compare policies with optimized parameter selections}
to obtain insights
into key cost/performance tradeoffs.
\revthree{We find that dynamic cache instantiation has the
potential to provide substantial cost
\revthree{reductions in some cases,}{reductions,}
but that this potential
is strongly dependent on the object popularity skew.
We also find that selective \emph{Cache on $k^{th}$ request}
cache insertion policies can be even more beneficial in this context
than with conventional edge caches.}{We find
  (among other things)
  that dynamic cache instantiation can provide substantial cost reductions,
  that
  potential reductions strongly dependent on the object popularity skew, and
  that
  selective cache insertion can be even more beneficial in this context than with conventional edge caches.}
\revExt{}{Finally, our contributions also include accurate and easy-to-compute
  approximations that are shown applicable to LRU caches under time-varying workloads.}}{Content-delivery applications
  can achieve scalability and reduce wide-area network traffic using geographically distributed caches.
  However, each deployed cache has an associated cost, and under time-varying request rates
  (e.g., a daily cycle) there may be long periods when the request rate from the local region
  is not high enough to justify this cost. Cloud computing offers a solution to problems of
  this kind, by supporting dynamic allocation and release of resources. In this paper,
  we analyze the potential benefits from dynamically instantiating caches using resources
  from cloud service providers. We develop novel analytic caching models that accommodate
  time-varying request rates, transient behavior as a cache fills following instantiation,
  and selective cache insertion policies. Within the context of a simple cost model,
  we then develop bounds and compare policies with optimized parameter selections to
  obtain insights into key cost/performance tradeoffs. We find that dynamic cache
  instantiation can provide substantial cost reductions, that potential reductions
  strongly dependent on the object popularity skew, and that selective cache insertion
  can be even more beneficial in this context than with conventional edge caches.
  Finally, our contributions also include accurate and easy-to-compute approximations
  that are shown applicable to LRU caches under time-varying workloads.}
\end{abstract}


\fancypagestyle{firststyle}
               {
                 \fancyhf{}
                 \fancyfoot[CF]{\tiny
                   \copyright IEEE (2021). This is the author's version of the work (as accepted). It is posted here by permission of IEEE for your personal use. Not for redistribution. 
                   \\The definitive version will be published in 
                   {\em  IEEE Transactions on Cloud Computing},
                   accepted Sept. 2021,
                   \url{https://doi.org/10.1109/TCC.2021.3115959}.}
               }
               \thispagestyle{firststyle}


%
%


%
%


\section{Introduction}

\revthree{Content delivery systems can improve performance and scalability
through use of geographically distributed caches that
serve their content to local client populations.
However, not all regions will always have sufficiently high
request rates to justify the cost of a local cache.
Of particular interest here, are cases where peak daily request
rates may be sufficient to justify a local cache,
whereas off-peak rates may not.
Substantial request rate variation according to a relatively
predictable daily cycle is commonly observed in content
delivery applications~\cite{GALM07,GAC+11,TFK+11}.}{The performance and scalability of content
  delivery systems benefit significantly from geographically distributed caches.  It is therefore
  not surprising that caching solutions for these systems have generated much research (Section~\ref{relatedwork}).
  However, despite the emergence of distributed, regional, and edge cloud computing offering
  a completely new service paradigm -- on-demand caching -- surprisingly few works have taken
  into account on-demand cache provisioning~\cite{CaNM19,SuKS16,DaCa14,CaLK14,KIM+17}, and, to our knowledge, no prior work
  has
  considered,
  rigorously modelled, and analyzed
  the problem of when to instantiate and release caches in such environments.}

\revthree{Ideally, we would like to incur the cost of
a cache only only for the portion of the day when
the request rate is sufficiently high to justify this cost.
\revrev{Cloud}{Distributed, regional, and edge cloud}
computing offers a potential solution to
problems of this kind by supporting the provisioning and
release of
resources on-demand.
However, to our knowledge, no prior work has investigated the
possible applicability of this paradigm to cache provisioning for
content delivery applications.}{We note that 
request rates of these systems typically differ between locations and
\revfour{vary over time (e.g., according to a relatively predictable daily cycle~\cite{GALM07,GAC+11,TFK+11}).}{the total aggregate request volumes for a given location typically vary over time according to a relatively predictable daily cycle~\cite{M20,FGL+20,GALM07,GAC+11,TFK+11}.  Such cycles have been reported for ISPs~\cite{FGL+20}, CDNs (e.g., Akamai on a per-country basis~\cite{M20}), university networks (e.g.,~\cite{GALM07,GAC+11}), enterprise networks (e.g.,~\cite{GAC+11}), and popular services (e.g., YouTube~\cite{TFK+11}).
\revDerek{Each of these example locations provides examples of places where regional caches often are used.}{These are also contexts where regional caches are often used.}}

\revfour{In}{Due to such daily patterns, in}
systems where the service provider pays on an on-demand basis, the cost of a local cache (in some locations) may therefore only be justified during the daily peak in the request rates.
  Ideally, we would like to incur the cost of a cache only when the request rate is sufficiently
  high to justify this cost.}

In this paper, we take a first look at the potential benefits from
dynamically instantiating and releasing
\revthree{caches.}{caches (e.g., based on daily cycles).}
\revthree{For this purpose,}{In particular,}
we develop
\revthree{}{novel}
analytic models of cache
\revthree{performance,}{performance that accommodate the important challenges of taking into account
  (i) arbitrarily time-varying request rates and
  (ii) periods of transient behavior when a cache fills following instantiation, and
  apply these models within the context of a simple cost model to study cost/performance tradeoffs.}
\revthree{Use of analytic modeling for this problem is challenging because
the models need to be able to accommodate arbitrarily
time-varying request rates, as well as
the period of transient behavior when a cache fills following instantiation.
Also, in addition to conventional \emph{Cache on $1^{st}$ request}
indiscriminate caching, selective cache insertion policies are
of interest, since dynamically instantiated caches may be relatively small,
and therefore cache pollution may be a particularly important concern.}{}
\revfour{}{Our \revDerek{model is}{models are} motivated by \revDerek{a system where there are}{scenarios in which a system has} many independently operated cache locations and each cache is dynamically allocated within an edge cloud environment. \revDerek{Thus, while}{Although} there are many geographically distributed caches, for the purpose of our analysis we \revDerek{consider}{can consider just} one such location.  \revDerek{However, in}{In} contrast to prior works we consider a novel cloud context in which the caches are dynamically instantiated so \revDerek{}{as} to minimize the delivery cost under time-varying request volumes.}

\revthree{To address these challenges,}{First, to accommodate time-varying request rates and periods of transient behavior,}
we develop a modelling approach based on what we term here
``request count window'' (RCW) caches.
Objects are evicted from an RCW cache if not requested over a window
consisting of the most recent $L$ requests,
where $L$ is a parameter of the
\revrev{caching system.}{system.}
As we show here empirically, similarly as with ``Time-to-Live'' (TTL)
caches~\cite{ChTW02, FrRR12, BDC+13, BGSC14, GaLM16} in scenarios with fixed request rates,
the performance of an RCW cache
closely approximates the performance of an LRU cache when the size of
the window (for an RCW cache,
measured in number of requests) is set such that the
average occupancy equals the
\revrev{size of the LRU cache.}{LRU cache size.}

\revthree{We}{Second, we} carry out analytic analyses of RCW caches for both
indiscriminate \emph{Cache on $1^{st}$ request} and selective
\emph{Cache on $k^{th}$ request} cache insertion policies.
\revrev{}{This includes the derivation of explicit, exact expressions for key cache performance metrics
  under the independent reference model,
  including (i) the hit and insertion rates for permanently allocated caches, and
  (ii) the average rates over the transient period during which a newly instantiated cache is filling.
  \revthree{We also derive approximate expressions of $\mathcal{O}(1)$ computational
cost for the cases of Zipf object popularities with parameter
$\alpha = 1$ and $\alpha = 0.5$.
These two cases are chosen as representative
of high and low popularity skew, respectively.}{We note that for this context,
    selective \emph{Cache on $k^{th}$ request} cache insertion policies are of particular interest,
    since dynamically instantiated caches may be relatively small,
    and therefore cache pollution may be a particularly important concern.}}
\revExt{}{We also derive approximate expressions of $\mathcal{O}(1)$ computational
  cost for the cases of Zipf object popularities with parameter
  $\alpha = 1$ and $\alpha = 0.5$.
  These two cases are chosen as representative
of high and low popularity skew, respectively.}
\revrev{}{Our RCW analysis makes no assumptions regarding inter-request time distributions or request rate variations,
  ensuring that our RCW results
  \revthree{}{(in contrast to prior TTL approximations~\cite{ChTW02, FrRR12, BDC+13, BGSC14, GaLM16})}
  can be used to approximate LRU cache performance under highly time-varying request volumes.
  In general, for time-varying workloads, the concept of RCW caches
  \revthree{also provides}{provides}
  a more natural choice than
  TTL caches when approximating fixed-capacity LRU caches.}

\revthree{In}{Third, in}
addition to the cache insertion policy, important design issues
in a dynamic cache instantiation system include the choice of cache 
size and the duration of the cache instantiation interval.
We develop optimization models for these parameters
for both \emph{Cache on $1^{st}$ request}
and \emph{Cache on $k^{th}$ request}.
We also develop bounds on the best potentially achievable
cost/performance
\revrev{tradeoffs.}{tradeoffs,
assessing how much room for improvement there may
be through use of more complex caching policies.}

\revfour{\revrev{We}{Finally, we}}{Fourth, we}
apply our analyses to obtain
insights into the potential cost reductions possible with
dynamic cache instantiation and explore key system tradeoffs.
\revDerek{
\revthree{We find that dynamic cache instantiation has the
potential to provide significant cost reductions in some cases,
but that this potential
is strongly dependent on the object popularity skew.}
{(1) We find that dynamic cache instantiation has the
  potential to provide significant cost
  reductions, sometimes more than halving the costs of (optimized) baselines
  that either use a cache or not, depending on which results in a lower cost.
  (2) The cost reductions are strongly dependent on the object popularity skew.}
When there is high skew, dynamic instantiation can work
\revthree{}{particularly}
well since a newly instantiated cache is
quickly populated with frequently requested items that will
capture a substantial fraction of the requests.
\revthree{}{(3)}
We also find that selective \emph{Cache on $k^{th}$ request}
cache insertion policies can be even more beneficial in this context
than with conventional edge caches, and that, when there is high
popularity skew, there is likely only modest room for improvement
in cost/performance
through use of more complex cache insertion and replacement policies.}{We find that:
\begin{itemize}
\item \revfour{Dynamic cache instantiation has the potential to provide significant cost reductions, sometimes more than halving the costs of (optimized) baselines that either use a cache or not, depending on which results in a lower cost.}{Dynamic cache instantiation has the potential to provide significant cost reductions, sometimes more than halving the costs of (optimized) baselines that uses a permanent cache, with the cache size selected so as to minimize the cost.}
\item The cost reductions are strongly dependent on the object popularity skew. When there is high skew, dynamic instantiation can work particularly well since a newly instantiated cache is
quickly populated with frequently requested items that will
capture a substantial fraction of the requests.
\item Selective \emph{Cache on $k^{th}$ request}
cache insertion policies can be even more beneficial in this context
than with conventional edge caches, and, when there is high
popularity skew, there is likely only modest room for improvement
in cost/performance
through use of more complex cache insertion and replacement policies.
\end{itemize}
Overall, these}
\revDerek{\revthree{These}{Overall, these}}{}
results
\revthree{suggest}{show}
that dynamic cache instantiation using \emph{Cache on $k^{th}$ request}
\revthree{may be}{is}
a promising approach for content delivery applications.

\revfour{}{Finally, it is important to note that there does not exist any analysis (from prior work) that captures the performance of LRU caches under time-varying workloads.  Our development of easy-to-compute approximation expressions of the performance of LRU caches under time varying workloads is therefore an important contribution.  The reason we use RCW for our analysis (rather than LRU) is in part because it enables both an exact analysis and because it provides a nice approximation for LRU caches, while still capturing the cache performance under time-varying workload volumes.}
\revfive{}{In contrast, exact analysis of large LRU caches is intractable.  Of course, in practice, we expect many systems to keep implementing LRU cache replacement policies or some variation thereof.}


\revthree{The remainder of the paper is organized as follows.
In Section~\ref{systemdescriptionandmetrics}, we describe our
workload and system assumptions,
and the caching policies and metrics we consider.}{{\bf Roadmap:}
  Section~\ref{systemdescriptionandmetrics} describes our workload and system assumptions,
the caching policies considered, and the metrics of interest.}
Section~\ref{alwaysonanalysis} presents our analysis of RCW caches for
the baseline case
\revthree{with no}{without}
use of dynamic instantiation.
\revthree{Dynamic instantiation is addressed in Section~\ref{transientperiodanalysis}, which}{Section~\ref{transientperiodanalysis}}
provides an analysis of the period of transient behavior as an RCW cache fills.
Optimization models and performance results for dynamic instantiation
are presented in Sections~\ref{optimizationmodel} and~\ref{comparisons},
respectively.
\revExt{}{Throughout the paper we derive
  and present results for both exact and $\mathcal{O}(1)$-approximations.}
Section~\ref{relatedwork} describes related work,
\revthree{and Section~\ref{conclusions} concludes with a summary
  and directions for future work.}{before Section~\ref{conclusions} concludes the paper.}

\section{System Description and Metrics}~\label{systemdescriptionandmetrics}
\revthree{\subsection{Workload Assumptions}\label{workloadassumptions}}{{\bf Workload Assumptions:}}
\revrev{We focus on a single region within the service area of a content delivery application.
  The baseline design that we consider does
  not use dynamic cache instantiation, instead using a permanently
  allocated (``always-on'') cache for this region.
  Of interest is the potential benefit from use of
  dynamic cache instantiation.  In Section~\ref{workloadassumptions},
  we describe our assumptions regarding the workload of content requests,
  while Sections~\ref{cachepolicies} and~\ref{metrics} describe
  the cache policies and system cost/performance metrics we consider,
  respectively.}{We focus on a single region within the service area of a content delivery application,
or a cache location to which a subset of geographically distributed clients are directed~\cite{ChST15}.}
\revrev{We}{For this cache location, we}
consider a time period of duration $T$
\revthree{(for example, one day),}{(e.g., one day),}
over which the total
\revrev{}{(aggregated over all objects)}
content request rate $\lambda(t)$ varies.
We assume that these variations are
predictable
\revthree{(for example, based on prior days),}{(e.g., based on prior days),}
and so for any desired cache instantiation duration $D < T$,
it would be possible to identify
in advance the interval of duration $D$ with
the highest average
request rate over all intervals of duration
$D$ within the time period.

Short-term temporal locality, non-stationary object popularities,
and high rates of new content creation make dynamic cache instantiation
potentially more promising, since they reduce the value of old
cache contents.
\revthree{A}{Here, we provide a}
conservative estimate of the benefits of dynamic cache
\revthree{instantiation can be achieved by}{instantiation,}
assuming a fixed set of objects with stationary object
popularities, and with
requests following the independent reference model.
We denote the number of objects by $N$, and index the objects
such that $p_i \geq p_{i+1}$ for $1 \leq i < N$, where $p_i$ denotes
the probability that a request is for
\revthree{object $i$. As special cases of object popularity skew, we consider both Zipf
with parameter $\alpha = 1$, and Zipf with $\alpha = 0.5$.
Commonly popularity skew is intermediate between these two cases.}{object $i$.}

\revthree{\subsection{Cache Policies}\label{cachepolicies}}{{\bf Cache Policies:}}
We model what we term here ``request count window'' (RCW) caches.
Objects are evicted from an RCW cache if not requested over a window
consisting of the most recent $L$ requests, where $L$ is a
\revthree{parameter of the caching system.}{system parameter.}
As we empirically demonstrate, the performance
of an RCW cache closely approximates the performance of an LRU cache
when the value of $L$ is set such that the average occupancy equals the
size of the LRU cache.

Both indiscriminate, \emph{Cache on $1^{st}$ request}, and
selective \emph{Cache on $k^{th}$ request}
cache insertion policies are considered.
\revthree{For integer $k > 1$, the \emph{Cache on $k^{th}$ request}
  insertion policy that we consider requires that}{For \emph{Cache on $k^{th}$ request} with $k$$>$$1$,
  we assume that}
the system
\revthree{maintain}{maintains}
some state information regarding
uncached objects that have been requested at least once
over a window consisting of
the most recent $W$ requests,
where $W$ is a policy parameter.
Specifically,
for each such ``caching candidate", a count of how many requests
are made for the object while it is a caching candidate is
maintained.
\revExt{Should this count reach $k$, the object is cached.
Should no request be made to the object for $W$ requests, the object
is removed as a caching candidate.
In either case,
if at some later point another request is made for the object
while uncached (i.e., after eviction from the cache if it had been
added),
the object becomes a caching candidate again, with count initialized
to one.}{When a request is
  received for an uncached object that was not already a caching
  candidate, the object becomes a caching candidate with count
  initialized to one. Should this count reach $k$, the object is
  cached. Should no request be made to the object for $W$
requests, the object is removed as a caching candidate.}

\revrev{}{For the dynamic instantiation, we assume that the cloud provider returns an empty cache when (re)instantiated.
This does not require us to make any assumption of the type of cache (e.g., in memory vs disk-based storage,
type of VMs, etc.).  However, we note that the cloud provider that is not able to rent out the resources to
serve other workloads may decide to only shut down disks/memory to save energy and in some of these cases
therefore potentially could return part of the cache in its original state.  For such a case, our analysis
provides a pessimistic performance bound.}

\revthree{\subsection{Metrics \revthree{}{and Cost Assumptions}}\label{metrics}}{{\bf Metrics and Cost Assumptions:}}
\revthree{The metrics of primary interest are the expected
fraction of requests over the entire time period
that are served locally from cache
and the cache cost.}{The metrics of primary interest are
  the expected {\em fraction of requests served locally} from cache (over the entire time period),
  and the {\em cache cost}.}
With dynamic cache instantiation,
\revthree{the expected fraction of requests served
\revrev{}{locally}
from the cache}{the first of these two metrics}
is given by
	$\frac{\bar{H}_{t_a : t_d} \int_{t_a}^{t_d} \lambda (t) \textrm{d}t}{\int_0^T \lambda (t) \textrm{d}t}$,
where
$t_a$ denotes the time at which the cache is allocated,
$t_d$ the time at which it is deallocated, and
$\bar{H}_{t_a : t_d}$
the average hit rate over this interval.
Note that the hit rate (probability) will vary
over the interval,
with the hit rate immediately after
\revthree{instantiation, for example,}{instantiation}
being zero
\revthree{since the cache is empty at this point.}{(empty cache).}

Implementations of dynamic cache instantiation could use a variety
of technologies.
One option would be to use dynamic allocation
of a virtual machine, with main memory used for the cache.
We assume here a simple cost model where the cost per unit time of
a cache of capacity $C$ objects is proportional
to $C + b$, where the constant $b$ captures
the portion of the cost that is independent of cache size.
The total cost over the period $T$
is then proportional to $(t_d - t_a) (C+b)$.
More complex cost models could be easily
\revthree{accommodated, however,}{accommodated;}
the only issue being the computational cost of evaluating the cost
function when solving our optimization models.

\revExt{\revthree{A secondary metric that we consider is the fraction of cache object
  insertion/retrieval operations that are insertions.}{In addition to the primary metrics, we also evaluate
the {\em fraction of cache insertions} (over all cache object insertion/retrieval operations).}
With dynamic cache instantiation,
this is given by $\bar{I}_{t_a : t_d}/(\bar{I}_{t_a : t_d} + \bar{H}_{t_a : t_d})$,
where $\bar{I}_{t_a : t_d}$
denotes the average cache insertion rate over the cache instantiation
\revthree{interval; i.e., the fraction of
  requests that result in the requested object being inserted into the cache.}{interval.}
\revrev{As with the hit rate, the insertion
  rate will vary over the duration of the instantiation interval. The}{The}
  fraction of cache operations that are insertions is an important
measure of overhead; cache insertions consume node resources, but do
not yield any benefit unless there are subsequent resulting cache hits.
\revrev{}{Finally, we use the
  \revthree{average number of objects $A$ in an RCW cache}{{\em average number of objects $A$ in an RCW cache}}
  to
  match the
  cache capacity $C$ of an LRU cache with similar performance.}}{In addition to the above metrics, when analyzing RCW
  caches we evaluate the hit rate $H$, the cache insertion rate
  $I$ (fraction of requests that result in object insertions into
  the cache) as well as the insertion fraction $I/(I + H)$, and
  the average number $A$ of objects in the cache. The insertion
  fraction is an important measure of overhead; cache insertions
  consume node resources, but do not yield any benefit unless
  there are subsequent resulting cache hits. The average number
  of objects in the cache is used to match the cache capacity $C$
of an LRU cache with similar performance.}

\section{RCW Cache Analysis}~\label{alwaysonanalysis}

In this section, we present analysis and performance results for
a permanently allocated RCW cache.
\revthree{Sections~\ref{cacheonfirst},~\ref{cacheonsecond}, and~\ref{cacheonkth}
consider the cases of indiscriminate caching, \emph{Cache on $2^{nd}$ request},
and \emph{Cache on $k^{th}$ request} for general $k \geq 2$, respectively.
\revthree{Table~\ref{notation} summarizes our notation.}{Table~\ref{notation} summarizes our notation.}}{}
\revfour{}{Section 3.1 derives exact expression for such an ``always on" RCW cache with arbitrary file object popularity.  Section 3.2 then derives approximations of $\mathcal{O}(1)$ computational cost for the case of Zipf object popularities with $\alpha = 1$ and $\alpha =0.5$.  Next, Section 3.3 leverages these results to derive $\mathcal{O}(1)$ approximation expressions for the RCW performance.  For readers wanting to skip the derivation details, we refer to equations (\ref{kobjs})-(\ref{kins}) or equations (\ref{kW=L}) for exact expressions, Table~\ref{tab:summary-approximations} for the key $\mathcal{O}(1)$ approximations, and to Section 3.4 for validation results and policy comparisons.}
\revExt{}{Table~\ref{notation} summarizes our notation.}

\subsection{\revthree{}{Exact ``Always on'' RCW Analysis}}

\revthree{\subsection{Cache on 1st Request}\label{cacheonfirst}}{{\bf Cache on 1$^{st}$ Request:}}
The probability that a request for object $i$ finds it in
the cache is given by $1 - (1 - p_i)^L$,
\revthree{since for a RCW cache using \emph{Cache on $1^{st}$ request},}{since}
object $i$ will be
in the cache if and only if at least one of the most recent $L$
requests was to object $i$.
The average number $A$ of objects in the cache, as seen by a random
request, the insertion rate $I$, and the hit rate $H$,
are therefore given by
{\footnotesize
\begin{gather}
    A \scalebox{0.75}[1.0]{\( = \)} N \scalebox{0.75}[1.0]{\( - \)} \sum_{i=1}^N (1 \scalebox{0.75}[1.0]{\( - \)} p_i)^L, ~
    I \scalebox{0.75}[1.0]{\( = \)} \sum_{i=1}^N p_i ( 1 \scalebox{0.75}[1.0]{\( - \)} p_i)^L, ~
    H \scalebox{0.75}[1.0]{\( = \)} 1 \scalebox{0.75}[1.0]{\( - \)} \sum_{i=1}^N p_i ( 1 \scalebox{0.75}[1.0]{\( - \)} p_i)^L. \label{1exact}
  \end{gather}}

\begin{table}
  \caption{Summary of notation}
  \label{notation}
  \vspace{-8pt}
{\small
  \begin{tabular}{|c|l|}
    \hline
        {\bf Notation} & {\bf Definition} \\
        \hline
        $T$ & Total duration of time period \\\hline
        $t_a$ & Time at which cache is allocated \\\hline
        $t_d$ & Time at which cache is deallocated \\\hline
        $N$ & Number of objects \\\hline
        $\alpha$ & Parameter of Zipf popularity distribution \\\hline
        $p_i$ & Probability that a request is to object $i$ \\\hline
        $L$ & Cache lifetime parameter (\# requests) \\\hline
        $W$ & \emph{Cache on $k^{th}$ request} window (\# requests) \\\hline
        $C$ & Cache capacity (\# objects)
        \\\hline
        $b$ & Cost per unit time independent of cache size\\\hline
        $H$ & Cache hit rate \\\hline
        $I$ & Cache insertion rate \\\hline
        $A$ & Average number of objects in cache \\\hline
        $\Theta_i$ & Object $i$ duration in cache (\# requests) \\\hline
        $\Delta_i$ & Object $i$ duration out of cache (\# requests) \\\hline
        $\gamma$ & Euler-Mascheroni constant ($\approx 0.577$) \\\hline
        $\Omega$ & Zipf normalization constant \\\hline
  \end{tabular}}
  \vspace{-10pt}
\end{table}

\revthree{\subsection{Cache on 2nd Request}\label{cacheonsecond}}{{\bf Cache on 2$^{nd}$ Request:}}
The expected value $E[\Theta_i]$ of the object $i$ duration in the cache,
measured in number of requests, is given by the average number of requests
until there is a sequence of $L$ requests in a row that do not include
a request for object $i$.  Since requests follow the independent
reference model and the probability of a request for object $i$
is $p_i$, this is the same as the average number
of flips of a biased coin that are required to get $L$ heads in a row,
with the probability of a head equal to $1 - p_i$:
{\footnotesize
\begin{equation}\label{Theta}
	E[\Theta_i] = \sum_{r=1}^L \frac{1}{(1-p_i)^r} = \frac{1 - (1 - p_i)^L }{p_i (1 - p_i)^L}.
\end{equation}}
With \emph{Cache on $2^{nd}$ request},
the expected value $E[\Delta_i]$ of the object $i$ duration out of the cache,
measured in number of requests, satisfies the following equation:
{\footnotesize
\begin{eqnarray}
  E[\Delta_i] & = & 1/p_i + (1 - p_i)^W ( W + E[\Delta_i] ) \nonumber\\
   & & + (1 - (1 - p_i)^W ) \left( 1 / p_i  - \frac{(1 - p_i)^W W}{1 - (1 - p_i)^W } \right)
\end{eqnarray}}
Here, the first term ($1/p_i$)
gives the expected number of requests until the first request
for object $i$ following its removal from the cache.
The second term gives the expected number of additional
requests until object $i$ is added to the cache, conditional on
the first request not being followed by another request within the window
$W$, multiplied by the probability of this condition.
The third term gives the expected number of additional
requests until object $i$ is added to the cache, conditional on
the first request being followed by another request within the window
$W$, multiplied by the probability of this condition.
Solving for $E[\Delta_i]$ yields:
{\footnotesize
\begin{equation}\label{k2Delta}
	E[\Delta_i] = \frac{2 - ( 1 - p_i)^W}{p_i (1 - (1 - p_i)^W )}.
\end{equation}}

\revthree{The average number $A$ of objects in the cache,
the cache hit rate $H$, and the cache
insertion rate $I$, are given by:}{Noting that
  $A = \sum_i \frac{E[\Theta_i]} {E[\Theta_i] + E[\Delta_i]}$,
$H = \sum_i p_i \frac{E[\Theta_i]} {E[\Theta_i] + E[\Delta_i]}$, and
  $I = \sum_i \frac{1} {E[\Theta_i] + E[\Delta_i]}$,
  we have:}
{\footnotesize
\begin{eqnarray}
  A = \sum_{i=1}^N \frac{1 - (1 - p_i)^L - (1 - p_i)^W + (1 - p_i)^{L+W}}{1 + (1 - p_i)^L - (1 - p_i)^W}, \label{2objs}\\
  H = \sum_{i=1}^N p_i \frac{1 - (1 - p_i)^L - (1 - p_i)^W + (1 - p_i)^{L+W}}{1 + (1 - p_i)^L - (1 - p_i)^W}, \label{2hit}\\
  I = \sum_{i=1}^N p_i \frac{(1 - p_i)^L - (1 - p_i)^{L+W}}{1 + (1 - p_i)^L - (1 - p_i)^W}. \label{2ins}
\end{eqnarray}}

\revthree{\subsection{Cache on kth Request}\label{cacheonkth}}{{\bf Cache on k$^{th}$ Request:}}
The analysis for general $k \geq 2$  differs from that for
\emph{Cache on $2^{nd}$ request} with respect to
$E[\Delta_i]$, the expected value of the object $i$ duration out of the cache,
measured in number of requests.
Denoting $E[\Delta_i]$
for \emph{Cache on $k^{th}$ request} by $E^k[\Delta_i]$, $E^k[\Delta_i]$
($k \geq 2$) can be expressed as a function of $E^{k-1}[\Delta_i]$ as follows:
{\footnotesize
  \begin{equation}
    E^k[\Delta_i] = \frac{E^{k-1}[\Delta_i] + (1 \scalebox{0.75}[1.0]{\( - \)} p_i)^W W +
      (1 \scalebox{0.75}[1.0]{\( - \)} (1 \scalebox{0.75}[1.0]{\( - \)} p_i)^W) \left( \frac{1}{p_i} \scalebox{0.75}[1.0]{\( - \)} \frac{(1 \scalebox{0.75}[1.0]{\( - \)} p_i)^W W}
      {1 \scalebox{0.75}[1.0]{\( - \)} (1 \scalebox{0.75}[1.0]{\( - \)} p_i)^W} \right)}{1 \scalebox{0.75}[1.0]{\( - \)} (1 \scalebox{0.75}[1.0]{\( - \)} p_i)^W},
\end{equation}}
with $E^1[\Delta_i]$ defined as $1 / p_i$.
The numerator of the right-hand side of this equation
gives the expected number of requests from when an object
is removed from cache or removed as a caching candidate, until it
next exits from the state in which it is a caching candidate with a count of
$k-1$ (either owing to being cached because
of a request occurring within the window $W$, or removed as a caching
candidate if no such request occurs).
The denominator is the probability of
being cached when exiting from the state in which it is a caching
candidate with a count of $k-1$,
and therefore
the inverse of the denominator
gives the expected number of times the object will enter this state
until it is finally cached.
Simplifying yields
{\footnotesize
\begin{equation}
	E^k[\Delta_i] = \frac{E^{k-1}[\Delta_i]}{1 - (1-p_i)^W} + \frac{1}{p_i},
\end{equation}}
implying
{\footnotesize
\begin{equation}
E^k[\Delta_i] =
\frac{1}{p_i} \left( \frac{1 - (1-(1-p_i)^W)^k} {(1-p_i)^W(1 - (1-p_i)^W)^{k-1}} \right).
\end{equation}}
Expressing $A$, $H$, and $I$ in terms of
$E^k[\Delta_i]$ and
\revthree{$E[\Theta_i]$ as in~(\ref{2objs}),~(\ref{2hit}), and~(\ref{2ins}),}{$E[\Theta_i]$,}
  and then substituting in the above expression for $E^k[\Delta_i]$
and the expression for $E[\Theta_i]$ from~(\ref{Theta}), yields:
{\footnotesize
\begin{eqnarray}
  A = \sum_{i=1}^N \frac{1 - (1 - p_i)^L}{1 - (1 - p_i)^L + \frac{(1 - p_i)^L (1 - (1 - (1 - p_i)^W)^k)}{(1 - p_i)^W (1 - (1 - p_i)^W)^{k-1}}}, \label{kobjs} \\
  H = \sum_{i=1}^N p_i \frac{1 - (1 - p_i)^L}{1 - (1 - p_i)^L + \frac{(1 - p_i)^L (1 - (1 - (1 - p_i)^W)^k)}{(1 - p_i)^W (1 - (1 - p_i)^W)^{k-1}}}, \label{khit} \\
  I = \sum_{i=1}^N p_i \frac{(1 - p_i)^L}{1 - (1 - p_i)^L + \frac{(1 - p_i)^L (1 - (1 - (1 - p_i)^W)^k)}{(1 - p_i)^W (1 - (1 - p_i)^W)^{k-1}}}. \label{kins}
\end{eqnarray}}
Note that for $W$=$L$,
equations~(\ref{kobjs}),~(\ref{khit}), and~(\ref{kins}) reduce to:
{\footnotesize
\begin{gather}
  A = \sum_{i=1}^N (1 - (1 - p_i)^L)^k, \quad
  H = \sum_{i=1}^N p_i (1 - (1 - p_i)^L)^k, \nonumber\\ 
  I = \sum_{i=1}^N p_i (1 - p_i)^L (1 - (1 - p_i)^L)^{k-1}.\label{kW=L}
\end{gather}}

\begin{table*}[t]
  \caption{\revExt{Example}{$\mathcal{O}(1)$} approximations.}
  \label{tab:summary-approximations}
  \vspace{-8pt}
         {\scriptsize
           \begin{tabular}{|c|c|p{6.6cm}|p{6.6cm}|}
             \cline{2-4}
             \multicolumn{1}{c|}{} & Policy (or sum) & Zipf, $\alpha=1$ & Zipf, $\alpha=0.5$ \\ \hline
             \multirow{3}{*}{\begin{sideways}{Sums}\end{sideways}}
             & $\sum_{i=1}^N (1 - p_i)^L$
             & {\tiny $N - \frac{L}{\ln N + \gamma} \biggl(\ln N - \ln \left(\frac{L}{\ln N + \gamma}\right) + 1 - \gamma + \frac{L}{2N(\ln N + \gamma)}\biggr)$}
             & $N - L + \frac{L^2}{4N} \left( \ln ((2N)/L) + \frac{L}{6N} + \frac{3}{2} - \gamma \right)$
             \\ \cline{2-4}
             & $\sum_{i=1}^N p_i (1 - p_i)^L$
             & $1 - \frac{\ln (L/(\ln N + \gamma)) + 2\gamma - L/(N(\ln N + \gamma))}{\ln N + \gamma}$
             & $1 - \frac{L}{2N} \left( \ln ((2N)/L) + \frac{L}{4N} + 1 - \gamma \right)$
             \\ \hline
             \multirow{3}{*}{\begin{sideways}{Always on (steady state)}\end{sideways}}
             & Cache on 1$^{st}$
             & {\tiny $\begin{array}{c}
                   A \approx \frac{L}{\ln N + \gamma} \left(\ln N - \ln \left(\frac{L}{\ln N + \gamma}\right) + 1 - \gamma + \frac{L}{2N(\ln N + \gamma)} \right) \\
                   I \approx 1 - \frac{\ln (L/(\ln N + \gamma)) + 2\gamma - L/(N(\ln N + \gamma))}{\ln N + \gamma} \\
                   H \approx \frac{\ln (L/(\ln N + \gamma)) + 2\gamma - L/(N(\ln N + \gamma))}{\ln N + \gamma}
                 \end{array}$}
             & $\begin{array}{c}
               A \approx L \left(1 - \frac{L}{4N} \left( \ln ((2N)/L) + \frac{L}{6N} + \frac{3}{2} - \gamma \right)\right) \\
               I  \approx 1 - \frac{L}{2N} \left( \ln ((2N)/L) + \frac{L}{4N} + 1 - \gamma \right) \\
               H \approx \frac{L}{2N} \left( \ln ((2N)/L) + \frac{L}{4N} + 1 - \gamma \right)
             \end{array}$
             \\ \cline{2-4}
             & Cache on 2$^{nd}$, $W=L$
             & $\begin{array}{c}
               A \approx \frac{(2 \ln 2 - L/(N(\ln N + \gamma))) L}{\ln N + \gamma} \\
               H \approx \frac{\ln (L / (\ln N + \gamma)) + 2 \gamma - \ln 2 }{\ln N + \gamma} \\
               I \approx \frac{\ln 2 - L/(N (\ln N + \gamma))}{\ln N + \gamma}
             \end{array}$
             & $\begin{array}{c}
               A \approx \frac{L^2}{2N} \left( \ln (N/(2L)) + \frac{L}{2N} + \frac{3}{2} - \gamma \right) \\
               H \approx \frac{L}{N} \left( \ln 2 - \frac{L}{4N} \right) \\
               I \approx \frac{L}{2N} \left( \ln (N/(2L)) + \frac{3L}{4N} + 1 - \gamma \right)
             \end{array}$
             \\ \cline{2-4}
             & Cache on k$^{th}$, $k \ge 3$, $W=L$
             & $\begin{array}{c}
               A \approx \frac{\left( \sum_{j=2}^k (-1)^j \binom{k}{j} j \ln(j) \right) L}{\ln N + \gamma} \\
               H \approx \frac{\ln (L/(\ln N + \gamma)) + 2\gamma - \sum_{j=2}^k (-1)^j \binom{k}{j} \ln(j)}{\ln N + \gamma} \\
               I \approx \frac{\sum_{j=2}^k (-1)^j \binom{k-1}{j-1} \ln(j)}{\ln N + \gamma}
             \end{array}$
             & {\tiny $\begin{array}{c}
                   A \approx \left\{ \begin{array}{ll}
                     \left( 9 \ln 3 - 12 \ln 2 - L/N \right) L^2 / (4N) & k = 3, \\
                     \left( \sum_{j=2}^k (-1)^{j+1} \binom{k}{j} j^2 \ln(j) \right) L^2 / (4 N) & k \geq 4. \\
                   \end{array}\right. \\
                   H \approx \frac{L}{2N} \sum_{j=2}^k (-1)^{j} \binom{k}{j} j \ln(j) \\
                   I \approx \left\{ \begin{array}{ll}
                     (L/(2N)) \left(3 \ln 3 - 4 \ln 2 - \frac{L}{2N} \right) & k = 3, \\
                     (L/(2N)) \sum_{j=1}^{k-1} (-1)^{j} \binom{k-1}{j} (j+1) \ln(j+1) & k \geq 4. \\
                   \end{array}\right.
                 \end{array}$}
             \\ \hline
             \multirow{3}{*}{\begin{sideways}{Transient}\end{sideways}}
             & Cache on 1$^{st}$
             & $\bar{H}_{\textrm{transient}} \approx \frac{\ln ( L/(\ln N + \gamma)) + 2\gamma - 1 - L/(2N(\ln N + \gamma))}{\ln N + \gamma}$
             & $\bar{H}_{\textrm{transient}} \approx \frac{L}{4N} \left( \ln ((2N)/L) + \frac{L}{6N} + \frac{3}{2} - \gamma \right)$
             \\ \cline{2-4}
             & Cache on 2$^{nd}$, $W=L$
             & $\bar{H}_{\textrm{transient}} \approx \frac{\ln (L/(\ln N + \gamma)) + 2\gamma - 2\ln 2}{\ln N + \gamma}$
             & $\bar{H}_{\textrm{transient}} \approx \frac{L}{N} \left( \ln 2 - \frac{L}{8N} - \frac{1}{4} \right)$
             \\ \cline{2-4}
             & Cache on k$^{th}$, $k \ge 3$, $W=L$
             & Insert above into $\bar{H}_{\textrm{transient}} = H + I -\frac{A}{L}$
             & Insert above into $\bar{H}_{\textrm{transient}} = H + I -\frac{A}{L}$
             \\ \hline
         \end{tabular}}
         \vspace{-12pt}
\end{table*}

\subsection{Summation approximations}\label{sec:summation-approx}

\revExt{}{We have derived approximate expressions
  of $\mathcal{O}(1)$ computational cost for the cases of
  Zipf object popularities with $\alpha$ = 1 and $\alpha$ = 0.5.
  Table~\ref{tab:summary-approximations} summarizes the key approximations that we obtain.
  As an important step in deriving these novel approximations,
  this subsection derives foundational summation approximations,
  which we then apply in Section~\ref{sec:rcw-approx} to derive the RCW approximations shown
  in Table~\ref{tab:summary-approximations}.
  Readers not interested in the derivations of these approximations
  can skip to the evaluation comparisons of RCW vs LRU
  and exact vs approximate RCW analysis provided in Section~\ref{sec:RCW-eval}.}

\subsubsection{Zipf with {\large $\alpha$} = 1}

Consider the case of a Zipf object popularity distribution with $\alpha = 1$,
and denote the normalization constant $\sum_{i=1}^N 1/i$
by $\Omega$.
Note that for large $N$, $\Omega \approx \ln N + \gamma$ where
$\gamma$ denotes the
Euler-Mascheroni constant ($\approx 0.577$).

For large $N$, $L$, and $g \to \infty$,
{\footnotesize
  \begin{align}\label{intapprox}
    \sum_{i=1}^N (1 - p_i)^L & \approx \sum_{i=1}^N e^{-L / (i \Omega)}
    & \approx \int_{L / (g (\ln N + \gamma)) }^N e^{-L / (x (\ln N + \gamma)) } \textrm{d}x.
\end{align}}
Using a Taylor series expansion for $e^y$ gives:
{\footnotesize
  \begin{align}\label{taylor}
    &\int_{L / (g (\ln N + \gamma)) }^N e^{-L / (x (\ln N + \gamma))} \textrm{d}x \nonumber \\
     & ~~~ = \int_{L / (g (\ln N + \gamma)) }^N \sum_{j=0}^{\infty} \frac{(-L / (x (\ln N + \gamma)))^j}{j!} \textrm{d}x \nonumber \\
    & ~~~ = N - \frac{L}{\ln N + \gamma} \left( \ln N + \frac{L}{2N (\ln N + \gamma)}
    - \sum_{j=2}^{\infty} \frac{(-L / (N (\ln N + \gamma)))^j}{(j+1)!j} \right) \nonumber \\
    & ~~~~~~- \frac{L}{\ln N + \gamma} \left( 1/g - \ln(L / (\ln N + \gamma)) + \ln(g) + \sum_{j=1}^{\infty} \frac{(-g)^j}{(j+1)!j} \right).
\end{align}}
Note that
{\footnotesize
\begin{equation}\label{Eieq}
  \ln (g) + \sum_{j=1}^{\infty} \frac{(-g)^j}{j!j} = \textrm{Ei}(-g) - \gamma,
\end{equation}}
where Ei is the exponential integral function, and that
{\footnotesize
\begin{equation}
  \sum_{j=1}^{\infty} \frac{(-g)^j}{(j+1)!j} - \sum_{j=1}^{\infty} \frac{(-g)^j}{j!j} = \frac{1}{g}\sum_{j=1}^{\infty} \frac{(-g)^{j+1}}{(j+1)!} =
  \frac{1}{g} \left( e^{-g} + g - 1 \right),
\end{equation}}
which tends to 1 as $g \to \infty$.
Also, for $g \to \infty$, Ei($-g$) $\to 0$.
Therefore, for $g \to \infty$,
{\footnotesize
\begin{equation}\label{geq}
  1/g + \ln(g) + \sum_{j=1}^{\infty} \frac{(-g)^j}{(j+1)!j} \to 1 - \gamma.
\end{equation}}
Substituting this result into~(\ref{taylor}),
and neglecting the terms in the summation on the
\revExt{second}{third}
line of~(\ref{taylor}) under the assumption
that $L$ is substantially smaller than $N (\ln N + \gamma)$, yields
{\footnotesize
  \begin{equation}\label{appendixoneminuspLapprox}
    \sum_{i=1}^N (1 - p_i)^L \approx N - \frac{L}{\ln N + \gamma} \biggl(\ln N - \ln \left(\frac{L}{\ln N + \gamma}\right)
    + 1 - \gamma + \frac{L}{2N(\ln N + \gamma)}\biggr).
\end{equation}}

Again assuming large $N$, $L$, and $g \to \infty$,
{\footnotesize
  \begin{equation}
    \sum_{i=1}^N p_i (1 - p_i)^L \approx \sum_{i=1}^N \frac{e^{-L / (i \Omega )}}{i \Omega }
    \approx \int_{L / (g (\ln N + \gamma))}^N \frac{e^{-L / (x (\ln N + \gamma))}}{x (\ln N + \gamma)} \textrm{d}x.
\end{equation}}
Using a Taylor series expansion for $e^y$ gives:
{\footnotesize
  \begin{align}\label{taylor2}
    &\int_{L / (g (\ln N + \gamma)) }^N \frac{e^{-L / (x (\ln N + \gamma))}}{x (\ln N + \gamma)} \textrm{d}x \nonumber \\
    & ~~ = \int_{L / (g (\ln N + \gamma))}^N \sum_{j=0}^{\infty} \frac{(-L)^j (x (\ln N + \gamma))^{-(j+1)}}{j!} \textrm{d}x \nonumber \\
    & ~~ = \frac{1}{\ln N + \gamma} \left( \ln(N) + \frac{L}{N ( \ln N + \gamma )}
    - \sum_{j=2}^{\infty} \frac{(-L / (N (\ln N + \gamma)))^j}{j!j} \right) \nonumber \\
    & ~~~~~ - \frac{1}{\ln N + \gamma} \left( \ln(L/ (\ln N + \gamma)) - \ln(g) - \sum_{j=1}^{\infty} \frac{(-g)^j}{j!j} \right).
\end{align}}
Applying~(\ref{Eieq}) and considering $g \to \infty$,
and neglecting the
terms in the summation on the second line
of~(\ref{taylor2}) yields
{\footnotesize
\begin{equation}\label{appendixponeminuspLapprox}
  \sum_{i=1}^N p_i ( 1 - p_i)^L \approx 1 - \frac{\ln (L/(\ln N + \gamma)) + 2\gamma - L/(N(\ln N + \gamma))}{\ln N + \gamma}.
\end{equation}}

\subsubsection{Zipf with {\large $\alpha$} = 0.5}

Consider
the case of a Zipf object popularity distribution with $\alpha = 0.5$,
and denote the normalization constant $\sum_{i=1}^N 1/\sqrt{i}$
by $\Omega$.
Note that for large $N$, $\Omega \approx 2 \sqrt{N}$.

For large $N$, $L$, and $g \to \infty$,
{\footnotesize
\begin{equation}
  \sum_{i=1}^N (1 - p_i)^L \approx \sum_{i=1}^N e^{-L / ((\sqrt{i}) \Omega)} \approx \int_{(L/ (2 g \sqrt{N}))^2}^N e^{-L / (2 \sqrt{x} \sqrt{N})} \textrm{d}x.
\end{equation}}
Using a Taylor series expansion for $e^y$ gives:
{\footnotesize
  \begin{align}\label{sqtaylor}
    &\int_{(L/ (2 g \sqrt{N} ))^2}^N e^{-L / (2 \sqrt{x} \sqrt{N})} \textrm{d}x \nonumber \\
    & ~~ = \int_{(L/ (2 g \sqrt{N}))^2}^N \sum_{j=0}^{\infty} \frac{(-L / (2 \sqrt{x} \sqrt{N}))^j}{j!} \textrm{d}x \nonumber \\
    & ~~ = N - L + (L^2/(8N)) \ln N + \frac{L^3}{24N^2} - \frac{L^2}{4N} \sum_{j=4}^{\infty} \frac{(-L/(2 N))^{j-2}}{j!(j/2 - 1)} \nonumber \\
    & ~~~~~~ - \frac{L^2}{4 N} \left( 1/g^2 - 2/g + \ln (L / (2 \sqrt{N})) - \ln(g) - \frac{2}{g^2} \sum_{j=3}^{\infty} \frac{(-g)^j}{j!(j-2)} \right).
\end{align}}
Applying~(\ref{Eieq}) as well as the Taylor series expansion for $e^y$,
note that
{\footnotesize
  \begin{align}
    &\frac{2}{g^2} \sum_{j=3}^{\infty} \frac{(-g)^j}{j!(j-2)} \nonumber \\
    & = \frac{1}{g^2} \sum_{j=3}^{\infty} \frac{(-g)^j}{(j-1)!(j-2)} +
    \frac{2}{g^2} \sum_{j=3}^{\infty} \frac{(-g)^j}{(j-1)!(j-2)} \left( \frac{1}{j} - \frac{1}{2} \right) \nonumber \\
    & = \frac{1}{g^2} \sum_{j=3}^{\infty} \frac{(-g)^j}{(j-1)!(j-2)}
    - \frac{1}{g^2} \sum_{j=3}^{\infty} \frac{(-g)^j}{j!} \nonumber \\
    & = \left( \frac{1}{g} \left( e^{-g} + g - 1 \right)
    + \textrm{Ei}(-g) - \gamma - \ln g \right)
    - \frac{1}{g^2} \left( e^{-g} - \frac{g^2}{2} + g - 1 \right).
\end{align}}
Therefore, for $g \to \infty$,
{\footnotesize
\begin{equation}
  1/g^2 - 2/g - \ln(g) - \frac{2}{g^2} \sum_{j=3}^{\infty} \frac{(-g)^j}{j!(j-2)} \to -\frac{3}{2} + \gamma.
\end{equation}}
Substituting this result into~(\ref{sqtaylor}), and neglecting
the terms in the summation
on the second
line of~(\ref{sqtaylor}) under
the assumption that $L$ is substantially smaller than $2N$, yields
{\footnotesize
  \begin{equation}\label{appendixsqoneminuspLapprox}
    \sum_{i=1}^N (1 - p_i)^L \approx N - L + \frac{L^2}{4N} \left( \ln ((2N)/L) + \frac{L}{6N} + \frac{3}{2} - \gamma \right).
\end{equation}}

Again assuming large $N$, $L$, and $g \to \infty$,
{\footnotesize
  \begin{equation}
    \sum_{i=1}^N p_i (1 - p_i)^L \approx \sum_{i=1}^N
    \frac{e^{-L / ((\sqrt{i}) \Omega)}}{(\sqrt{i}) \Omega}
    \approx \int_{(L/ (2 g \sqrt{N} ))^2}^N
    \frac{e^{-L / (2 \sqrt{x} \sqrt{N})}}{2 \sqrt{x} \sqrt{N}} \textrm{d}x.
\end{equation}}
Using a Taylor series expansion for $e^y$ gives:
{\footnotesize
  \begin{align}\label{sqtaylor2}
    &\int_{(L/ (2 g \sqrt{N} ))^2}^N \frac{e^{-L / (2 \sqrt{x} \sqrt{N})}}{2 \sqrt{x} \sqrt{N}} \textrm{d}x
    = \int_{(L/ (2 g \sqrt{N}))^2}^N \sum_{j=0}^{\infty} \frac{(-L)^j}{j!(2 \sqrt{x} \sqrt{N})^{j+1}} \textrm{d}x \nonumber \\
    &= 1 - (L/(4N)) \ln N - L^2/(8N^2) - \sum_{j=3}^{\infty} \frac{(-L/(2 N))^j}{j!(j - 1)} \nonumber \\
    &- \frac{L}{2 N} \left( 1/g - \ln (L/(2 \sqrt{N})) + \ln(g) + \sum_{j=2}^{\infty} \frac{(-g)^{j-1}}{j!(j-1)} \right).
\end{align}}
Finally, making the assumption that $L$ is substantially smaller than $2N$,
we neglect the terms in the summation
on the second line of~(\ref{sqtaylor2}).
Applying~(\ref{geq}) then yields
{\footnotesize
\begin{equation}\label{appendixsqponeminuspLapprox}
  \sum_{i=1}^N p_i (1 - p_i)^L \approx 1 - \frac{L}{2N} \left( \ln ((2N)/L) + \frac{L}{4N} + 1 - \gamma \right).
\end{equation}}

\subsection{RCW Cache Approximations}\label{sec:rcw-approx}


\subsubsection{Cache on 1st Request, Zipf with {\large $\alpha$} = 1}

Consider now
the case of a Zipf popularity distribution with $\alpha = 1$.
\revExt{Assuming large $N$ and $L$, such that $L$ is substantially smaller
  than $N (\ln N + \gamma)$
  where $\gamma$ denotes the Euler-Mascheroni constant ($\approx 0.577$),
  \revExt{in the Appendix}{in Section~\ref{sec:summation-approx}}
  we derive the following approximations
  for $\sum_{i=1}^N ( 1 - p_i)^L$ and $\sum_{i=1}^N p_i ( 1 - p_i)^L$:
  {\footnotesize
  \begin{equation}\label{oneminuspLapprox}
    \sum_{i=1}^N (1 - p_i)^L \approx N - \frac{L}{\ln N + \gamma} \biggl(\ln N - \ln \left(\frac{L}{\ln N + \gamma}\right)
    + 1 - \gamma + \frac{L}{2N(\ln N + \gamma)}\biggr),
  \end{equation}}
  {\footnotesize
  \begin{equation}\label{poneminuspLapprox}
    \sum_{i=1}^N p_i ( 1 - p_i)^L \approx 1 - \frac{\ln (L/(\ln N + \gamma)) + 2\gamma - L/(N(\ln N + \gamma))}{\ln N + \gamma}.
  \end{equation}}
  Applying~(\ref{oneminuspLapprox}) to the equation for $A$ in~(\ref{1exact}),
  and~(\ref{poneminuspLapprox})}{Applying~(\ref{appendixoneminuspLapprox}) to the equation for $A$ in~(\ref{1exact}),
  and~(\ref{appendixponeminuspLapprox})}
to the equations for $I$ and $H$, yields:
{\footnotesize
  \begin{equation}\label{1objs}
    A \approx \frac{L}{\ln N + \gamma} \left(\ln N - \ln \left(\frac{L}{\ln N + \gamma}\right) + 1 - \gamma
    + \frac{L}{2N(\ln N + \gamma)} \right),
\end{equation}}
{\footnotesize
\begin{equation}\label{1insr}
  I \approx 1 - \frac{\ln (L/(\ln N + \gamma)) + 2\gamma - L/(N(\ln N + \gamma))}{\ln N + \gamma},
\end{equation}}
{\footnotesize
\begin{equation}\label{1hit}
  H \approx \frac{\ln (L/(\ln N + \gamma)) + 2\gamma - L/(N(\ln N + \gamma))}{\ln N + \gamma}.
\end{equation}}

As we show empirically,
the performance of an RCW cache
closely approximates the performance of an LRU cache when $L$
is set such that the
average occupancy equals the size of the LRU cache.
Suppose that the LRU cache capacity $C = N^\beta$ for $0 < \beta < 1$.
Equating the LRU cache capacity to the approximation for
$A$ given in~(\ref{1objs}) yields
{\footnotesize
  \begin{equation}\label{Leq}
    N^\beta = \frac{L}{\ln N + \gamma} \left(\ln N - \ln \left(\frac{L}{\ln N + \gamma}\right) + 1 - \gamma + \frac{L}{2N(\ln N + \gamma)} \right).
\end{equation}}
An accurate approximation for the value of $L$ satisfying this
equation in the region of interest can be obtained by substituting for
$L$ in this equation with $N^\beta/((1-(\ln N/(\ln N + \gamma)) \beta)(1+a))$,
using the approximation
$\ln(1 + a) \approx a$ when $\left|a\right| < 1$,
neglecting the last term on the right-hand
side, and then solving for $a$ to obtain:
{\footnotesize
\begin{equation}
  a = \frac{\ln ((1-\beta) \ln N + \gamma) + 1 - 2 \gamma }{( 1 - \beta) \ln N - (1 - \gamma )}.
\end{equation}}
This yields:
{\footnotesize
  \begin{equation}\label{1L}
    L \approx
    \frac{N^\beta (\ln N + \gamma)((1-\beta) \ln N + \gamma - 1)} {((1- \beta) \ln N + \gamma)((1-\beta) \ln N - \gamma + \ln ((1-\beta) \ln N + \gamma))}.
\end{equation}}
Note that for large $N$, $L$ is substantially smaller than
$N (\ln N + \gamma)$, as
was assumed for the
\revExt{approximations~(\ref{oneminuspLapprox}) and~(\ref{poneminuspLapprox}).}{approximations~(\ref{appendixoneminuspLapprox})
  and~(\ref{appendixponeminuspLapprox}).}
Substituting into expressions~(\ref{1hit}) and~(\ref{1insr}) yield
cache hit rate and corresponding insertion rate approximations.
For the hit rate,
the resulting approximation is
$\beta$ minus a term that (slowly) goes to zero as $N \to \infty$:
{\footnotesize
\begin{align}
  &H \approx \beta - \Biggl(
  \frac{\ln \left(\frac{((1- \beta) \ln N + \gamma)((1-\beta) \ln N - \gamma + \ln ((1-\beta) \ln N + \gamma))}{(1-\beta)\ln N + \gamma - 1} \right)} {\ln N + \gamma} + \nonumber \\
       &\frac{ \frac{(1 -\beta) \ln N + \gamma - 1}{N^{1-\beta} ((1- \beta) \ln N + \gamma)((1-\beta) \ln N - \gamma + \ln ((1-\beta) \ln N + \gamma))} - \gamma (2 - \beta) }{\ln N + \gamma} \Biggr).
\end{align}}
We observe that further approximations can yield a
simpler approximation for $H$, accurate over a broad range of
cache sizes, of $\beta - c(1-\beta)/(2-\beta)$ where $c$ is
a small constant dependent on $N$ (e.g. $c = 1/3$ gives good results
for $N$ in the 10,000 to 100,000 range).
In contrast, note that the hit rate when the cache is kept filled with
the $\lfloor{C}\rfloor$
most popular objects (the optimal policy under the IRM assumption,
without knowledge of future requests) is given in this case
by $\sum_{i=1}^{\lfloor{C}\rfloor} p_i \approx (\ln C + \gamma ) / (\ln N + \gamma )$.
When $C = N^{\beta}$, this equals $\beta + \gamma (1 - \beta )/(\ln N + \gamma )$.

\subsubsection{Cache on 1st Request, Zipf with {\large $\alpha$} = 0.5}

Consider now the case of a Zipf popularity distribution with
$\alpha = 0.5$.
\revExt{Assuming large $N$ and $L$ such
  that $L$ is substantially smaller than $ 2 N$,
  \revExt{in the Appendix}{in Section~\ref{sec:summation-approx}}
  we derive the following approximations
  for $\sum_{i=1}^N ( 1 - p_i)^L$ and $\sum_{i=1}^N p_i ( 1 - p_i)^L$:
  {\footnotesize
  \begin{align}\label{sqoneminuspLapprox}
    \sum_{i=1}^N (1 - p_i)^L \approx N - L + \frac{L^2}{4N} \left( \ln ((2N)/L) + \frac{L}{6N} + \frac{3}{2} - \gamma \right),
  \end{align}}
  {\footnotesize
  \begin{equation}\label{sqponeminuspLapprox}
    \sum_{i=1}^N p_i (1 - p_i)^L \approx 1 - \frac{L}{2N} \left( \ln ((2N)/L) + \frac{L}{4N} + 1 - \gamma \right).
  \end{equation}}
  Applying~(\ref{sqoneminuspLapprox})
  to the equation for $A$ in~(\ref{1exact}), and~(\ref{sqponeminuspLapprox})}{Applying~(\ref{appendixsqoneminuspLapprox})
  to the equation for $A$ in~(\ref{1exact}), and~(\ref{appendixsqponeminuspLapprox})}
to the equations for $I$ and $H$, yields:
{\footnotesize
\begin{equation}\label{sq1objs}
  A \approx L \left(1 - \frac{L}{4N} \left( \ln ((2N)/L) + \frac{L}{6N} + \frac{3}{2} - \gamma \right)\right),
\end{equation}}
{\footnotesize
\begin{equation}\label{sq1insr}
  I  \approx 1 - \frac{L}{2N} \left( \ln ((2N)/L) + \frac{L}{4N} + 1 - \gamma \right),
\end{equation}}
{\footnotesize
\begin{equation}\label{sq1hit}
  H \approx \frac{L}{2N} \left( \ln ((2N)/L) + \frac{L}{4N} + 1 - \gamma \right).
\end{equation}}

Suppose now that the corresponding LRU
cache capacity $C = f N$ for some $f > 0$.
Equating the cache capacity $C$ to the approximation for the average number
of objects in the cache as given by~(\ref{sq1objs}) gives
{\footnotesize
\begin{equation}\label{sqLeq}
  fN = L \left(1 - \frac{L}{4N} \left( \ln ((2N)/L) + \frac{L}{6N} + \frac{3}{2} - \gamma \right)\right).
\end{equation}}
An accurate approximation for the value of $L$ satisfying this equation
for $L \leq 1.5 N$ can be obtained by
writing $L$ as $fN/(1+a)$, using $\ln (1 + a) \approx a$,
and neglecting the
$L/(6N)$ term, yielding the following equation for $a$:
{\footnotesize
\begin{equation}
  a^2 + \left(1 + \frac{f}{4} \right) a + \frac{f}{4} \left(\ln(2/f) + \frac{3}{2} - \gamma \right) = 0.
\end{equation}}
Solving for $a$ gives
{\footnotesize
\begin{equation}\label{sq1L}
  L \approx \frac{2 f N}{1 - f/4 + \sqrt{(1 + f/4)^2 - f(\ln(2/f) + 3/2 - \gamma )}}.
\end{equation}}
The relation $L \leq 1.5 N$ corresponds to an upper bound on
$f$ of about 0.68.
Substitution into~(\ref{sq1insr}) and~(\ref{sq1hit}) yields approximations
for the insertion and hit rates, respectively.
For small/moderate $f$ (e.g., $f \leq 0.2$,
so that the cache capacity is at most 20\% of the
objects), a simpler approximation for $H$ is $(f/2) \ln (c/f)$ where $c$
is a suitable constant such as 4.5.
Note the considerable contrast between the scaling of hit rate with
cache size for $\alpha = 1$ versus
$\alpha = 0.5$.
Also, when $\alpha = 0.5$ there is a bigger gap with respect to the
hit rate when the cache is kept filled with
the most popular objects.
In this case, the hit rate is
$\sum_{i=1}^{\lfloor{C}\rfloor} p_i \approx \sqrt{C}/ \sqrt{N}$.
When $C = f N$, this equals $\sqrt{f}$.


\subsubsection{Cache on 2nd Request, W=L, Zipf with {\large $\alpha$} = 1}

Consider now the case of $W=L$, and a Zipf object popularity distribution with $\alpha = 1$.
\revExt{Applying~(\ref{oneminuspLapprox}) to~(\ref{2objs}), and~(\ref{poneminuspLapprox})}{Applying~(\ref{appendixoneminuspLapprox})
  to~(\ref{2objs}), and~(\ref{appendixponeminuspLapprox})}
to~(\ref{2hit}) and~(\ref{2ins}),
yields
{\footnotesize
\begin{equation}\label{2approx-A}
  A \approx \frac{(2 \ln 2 - L/(N(\ln N + \gamma))) L}{\ln N + \gamma},
\end{equation}}
{\footnotesize
\begin{equation}\label{2approx-H}
  H \approx \frac{\ln (L / (\ln N + \gamma)) + 2 \gamma - \ln 2 }{\ln N + \gamma}
\end{equation}}
{\footnotesize
\begin{equation}\label{2approx-I}
  I \approx \frac{\ln 2 - L/(N (\ln N + \gamma))}{\ln N + \gamma}.
\end{equation}}
With respect to the range of values for $L$
for which these approximations are accurate,
note that, when $W=L$,~(\ref{2objs}),~(\ref{2hit}), and~(\ref{2ins})
include both $(1-p_i)^L$ and $(1-p_i)^{2L}$ terms.
Therefore, when $L$ is substantially smaller than $N (\ln N + \gamma)$,
but $2L$ is not, the accuracy of these approximations
is uncertain \emph{a priori}, and requires experimental
assessment.
A similar issue arises in the case of $\alpha = 0.5$, and
for \emph{Cache on $k^{th}$ request} with $k > 2$.

Equating the corresponding LRU cache capacity $C$
to the approximation for the average
number $A$ of objects in the cache as given in~(\ref{2approx-A}),
solving for $L$, and then applying the approximation $\sqrt{1-x} \approx
1 - x/2 - x^2/8$ for small $x$ yields:
{\footnotesize
\begin{equation}\label{2L}
  L \approx \frac{C (\ln N + \gamma) (1+ C/(4(\ln 2)^2 N) )}{2\ln 2}.
\end{equation}}
If the cache capacity $C = N^{\beta}$ for $0 < \beta < 1$,
substituting from~(\ref{2L}) into the
expression for $H$ in~(\ref{2approx-H}) yields an approximation
for the cache hit rate which for large $N$ is very close to $\beta$:
{\footnotesize
\begin{equation}
  H \approx \beta - \frac{\ln (4 \ln 2) - \ln (1+ 1/(4(\ln 2)^2 N^{1-\beta})) - \gamma (2 - \beta )}{\ln N + \gamma},
\end{equation}}
while substitution into the expression for $I$ in~(\ref{2approx-I}) yields
an approximation for the cache insertion rate.

\subsubsection{Cache on 2nd Request, W=L, Zipf with {\large $\alpha$} = 0.5}

For the case of $W = L$ and a Zipf object popularity distribution
with $\alpha = 0.5$,
\revExt{applying~(\ref{sqoneminuspLapprox}) to~(\ref{2objs}), and~(\ref{sqponeminuspLapprox})}{applying~(\ref{appendixsqoneminuspLapprox}) to~(\ref{2objs}), and~(\ref{appendixsqponeminuspLapprox})}
to~(\ref{2hit}) and~(\ref{2ins}), yields:
{\footnotesize
\begin{equation}\label{sq2-A}
  A \approx \frac{L^2}{2N} \left( \ln (N/(2L)) + \frac{L}{2N} + \frac{3}{2} - \gamma \right),~~~~
\end{equation}}
{\footnotesize
\begin{equation}\label{sq2-H}
  H \approx \frac{L}{N} \left( \ln 2 - \frac{L}{4N} \right),~~~~
\end{equation}}
{\footnotesize
\begin{equation}\label{sq2-I}
  I \approx \frac{L}{2N} \left( \ln (N/(2L)) + \frac{3L}{4N} + 1 - \gamma \right).
\end{equation}}

Suppose now that the corresponding LRU
cache capacity $C = fN$ for some $f > 0$.
Equating the cache capacity
$C$ to the approximation for the average number $A$ of objects in the
cache as given in~(\ref{sq2-A}),
writing $L$ as $f^{\frac{1}{2}}N/(1+a)$, and employing the approximations
$\ln (1+a) \approx a$ and $1/(1+a) \approx 1-a$
yields the following equation for $a$:
{\footnotesize
\begin{equation}
  2a^2 + \left(3 + \frac{f^{\frac{1}{2}}}{2} \right) a + \ln(2f^{\frac{1}{2}}) - \frac{f^{\frac{1}{2}}}{2} + \frac{1}{2} + \gamma = 0.
\end{equation}}
Solving for $a$ gives
{\footnotesize
\begin{equation}\label{sq2L}
  L \approx \frac{4 f^{\frac{1}{2}} N}{1 - \frac{f^{\frac{1}{2}}}{2} + \sqrt{\left(3 + \frac{f^{\frac{1}{2}}}{2}\right)^2 - 8(\ln(2f^{\frac{1}{2}}) - \frac{f^{\frac{1}{2}}}{2} + \frac{1}{2} + \gamma )}}.
\end{equation}}
Substitution into the expressions for $H$ and $I$ in~(\ref{sq2-H}) and~(\ref{sq2-I})
yields approximations for the hit and insertion rates.
For small/moderate $f$,
a rough approximation for $H$ is $c f^{1/2}$ where $c$
is a suitable constant such as 0.7.


\subsubsection{Cache on kth Request, W=L, Zipf with {\large $\alpha$} = 1}

Consider now the case of $W=L$,
and a Zipf object popularity distribution with $\alpha = 1$.
Writing out $(1 - (1 - p_i)^L)^k$ as a
\revrev{power series}{polynomial}
in $(1 - p_i)^L$
and
\revExt{applying~(\ref{oneminuspLapprox})}{applying~(\ref{appendixoneminuspLapprox})}
yields, for $k \geq 3$,
the following approximation for
the average number of objects in the cache:
{\footnotesize
\begin{equation}\label{kobjsapprox}
  A \approx \frac{\left( \sum_{j=2}^k (-1)^j \binom{k}{j} j \ln(j) \right) L}{\ln N + \gamma}.
\end{equation}}
\revExt{Applying~(\ref{poneminuspLapprox})}{Applying~(\ref{appendixponeminuspLapprox})}
the cache hit rate
can be approximated by
{\footnotesize
\begin{equation}\label{khitapprox}
  H \approx \frac{\ln (L/(\ln N + \gamma)) + 2\gamma - \sum_{j=2}^k (-1)^j \binom{k}{j} \ln(j)}{\ln N + \gamma},
\end{equation}}
and, for $k \geq 3$, the cache insertion rate by
{\footnotesize
\begin{equation}\label{kinsapprox}
  I \approx \frac{\sum_{j=2}^k (-1)^j \binom{k-1}{j-1} \ln(j)}{\ln N + \gamma}.
\end{equation}}

\begin{figure*}[t]
  \centering
  \subfigure[$k=1$]{
    \includegraphics[trim = 0mm 6mm 0mm 0mm, width=0.30\textwidth]{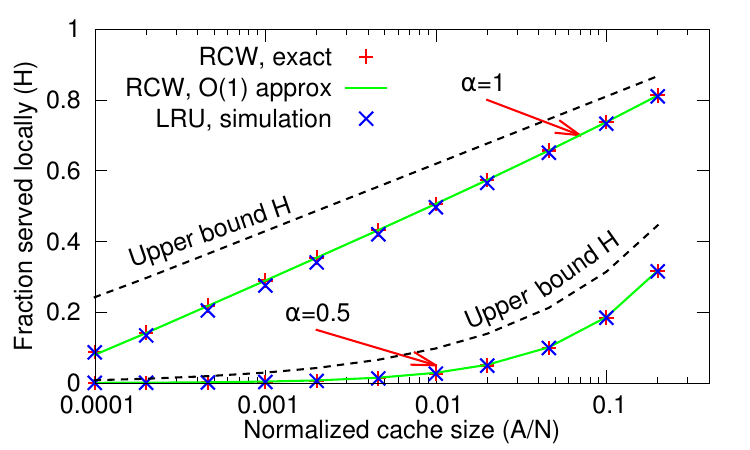}}
  \hspace{-10pt}
  \subfigure[$k=2$]{
    \includegraphics[trim = 0mm 6mm 0mm 0mm, width=0.30\textwidth]{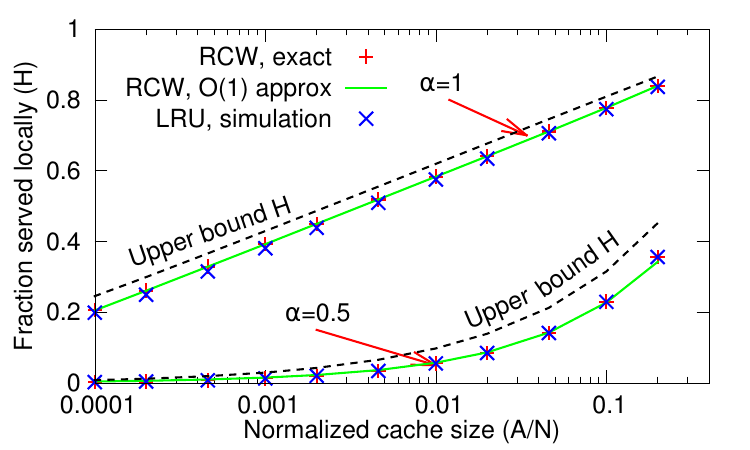}}
  \hspace{-10pt}
  \subfigure[$k=4$]{
    \includegraphics[trim = 0mm 6mm 0mm 0mm, width=0.30\textwidth]{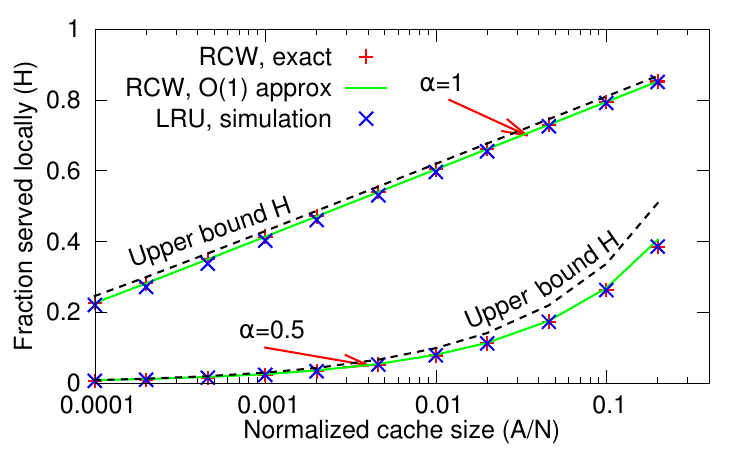}}
  \vspace{-12pt}
  \caption{Performance of \emph{Cache on $k^{th}$ request} ($N$$=$$100,000$); dashed lines show 
  hit rate when 
  cache is kept filled with the $\lfloor{C}\rfloor$ most popular objects.}
  \label{fig:accuracy-k124}
  \vspace{-12pt}
\end{figure*}

Equating the corresponding LRU
cache capacity $C$ to the approximation for the average
number of objects in the cache as given by expression~(\ref{kobjsapprox}),
and solving for $L$, yields, for $k \geq 3$,
{\footnotesize
\begin{equation}\label{kL}
  L \approx \frac{C (\ln N + \gamma)}{\sum_{j=2}^k (-1)^j \binom{k}{j} j \ln(j)}.
\end{equation}}
If the cache capacity $C = N^\beta$ for $0 < \beta < 1$,
substituting from~(\ref{kL}) into
expression~(\ref{khitapprox})
yields, for $k \geq 3$, an approximation for the cache hit rate which
for large $N$ is very close to $\beta$,
while the cache insertion rate
can be approximated using expression~(\ref{kinsapprox}).

\subsubsection{Cache on kth Request, W=L, Zipf with {\large $\alpha$} = 0.5}

For the case of $W = L$ and a Zipf object popularity distribution
with $\alpha = 0.5$,
\revExt{applying~(\ref{sqoneminuspLapprox})}{applying~(\ref{appendixsqoneminuspLapprox})}
to the equation for $A$ in~(\ref{kW=L})
yields:
{\footnotesize
\begin{equation}\label{sqkobjs}
  A \approx \left\{ \begin{array}{ll}
    \left( 9 \ln 3 - 12 \ln 2 - L/N \right) L^2 / (4N) & k = 3, \\
    \left( \sum_{j=2}^k (-1)^{j+1} \binom{k}{j} j^2 \ln(j) \right) L^2 / (4 N) & k \geq 4. \\
  \end{array}\right.
\end{equation}}
\revExt{Applying~(\ref{sqponeminuspLapprox})}{Applying~(\ref{appendixsqponeminuspLapprox})}
to the equation for $H$ in~(\ref{kW=L})
yields
the following approximation for the cache hit rate for $k \geq 3$:
{\footnotesize
\begin{equation}\label{sqkhitr}
  H \approx \frac{L}{2N} \sum_{j=2}^k (-1)^{j} \binom{k}{j} j \ln(j).
\end{equation}}
\revExt{Applying~(\ref{sqponeminuspLapprox})}{Applying~(\ref{appendixsqponeminuspLapprox})}
to the equation for $I$ in~(\ref{kW=L}) yields:
{\footnotesize
\begin{equation}\label{sqkins}
  I \approx \left\{ \begin{array}{ll}
    (L/(2N)) \left(3 \ln 3 - 4 \ln 2 - \frac{L}{2N} \right) & k = 3, \\
    (L/(2N)) \sum_{j=1}^{k-1} (-1)^{j} \binom{k-1}{j} (j+1) \ln(j+1) & k \geq 4. \\
  \end{array}\right.
\end{equation}}

Suppose now that the corresponding LRU
cache capacity $C = fN$ for some $f > 0$.
Equating the cache capacity
$C$ to the approximation for the average number of objects in
the cache given in~(\ref{sqkobjs}) for $k \geq 4$, and solving for $L$,
yields
{\footnotesize
\begin{equation}\label{sqkL}
  L \approx \frac{2 f^{\frac{1}{2}} N}{\sum_{j=2}^k (-1)^{j+1} \binom{k}{j} j^2 \ln(j)}.
\end{equation}}
Substitution into~(\ref{sqkins}) ($k \geq 4$ case) and~(\ref{sqkhitr})
yields approximations for the insertion and hit rates.
Note that the approximation for $H$ simplifies in this case
to $c f^{1/2}$ where $c$ is a $k$-dependent constant.

Equating the corresponding LRU cache capacity $C = fN$ to the approximation
for the average number of objects in the cache given in~(\ref{sqkobjs})
for $k=3$,
writing $L$ as $2 f^{\frac{1}{2}}N/(1+a)$, and employing the approximation
$1/(1+a) \approx 1-a+a^2$
for the $L/N$ term,
yields the following equation for $a$:
{\footnotesize
\begin{equation}
  (1 + 2f^{\frac{1}{2}}) a^2 + 2 \left(1 - f^{\frac{1}{2}} \right) a +
  1 + 2 f^{\frac{1}{2}} - 9 \ln 3 + 12 \ln 2 = 0.
\end{equation}}
Solving for $a$ gives, for $k=3$,
{\footnotesize
\begin{equation}\label{sq3L}
  L \approx \frac{2 f^{\frac{1}{2}} (1+2 f^{\frac{1}{2}}) N }{3 f^{\frac{1}{2}} + \sqrt{(f^{\frac{1}{2}} -1)^2 - (1+2 f^{\frac{1}{2}})(1+ 2 f^{\frac{1}{2}} - 9 \ln 3 + 12 \ln 2)}}.
\end{equation}}
Substitution into~(\ref{sqkins}) ($k=3$ case) and~(\ref{sqkhitr}) yield
approximations for the cache insertion and hit rates, respectively.
As with \emph{Cache on $2^{nd}$ request},
a rough approximation for $H$ is $c f^{1/2}$ for a constant $c$.


\subsection{Validation and Performance Results}\label{sec:RCW-eval}

\revfour{}{We have used simulations to validate the exact RCW models.  However, since the simulation results for RCW (``RCW sim") and the exact analytic results (``RCW exact") end up being the same (and the simulations just validate the exact expressions) we only show one of the two in the following figures.  Instead, these following figures focus on the comparisons between the simulated values of LRU (i.e., ``LRU sim") and the exact (or simulated) RCW values (again ``RCW exact” has the same values as ``RCW sim”) and $\mathcal{O}(1)$ approximations (i.e., ``RCW approx.").}

\revthree{}{Figure~\ref{fig:accuracy-k124} compares our exact RCW cache hit rate results (red '+' markers), using $W$$=$$L$, with the results from simulations of corresponding fixed-capacity LRU caches (blue '$\times$' markers),
  for $N$$=$$100,000$ objects, different $k$, and over a large range of cache sizes ($A/N$$=$$0.0001$ corresponds here to $A$$=$$10$).
  Also shown in the figure are the {\em upper bound}
  hit rate (dashed black lines), corresponding to when the cache is kept filled with the $\lfloor{C}\rfloor$ most popular objects,
  and $\mathcal{O}(1)$ {\em approximations} (green lines)
  derived in Section~\ref{sec:rcw-approx} for the special cases of Zipf distributions
  with  $\alpha$$=$$1$ and $\alpha$$=$$0.5$.
We
note that popularity skew typically is intermediate between these two cases.}

\revthree{}{For the simulations, we set the LRU cache size $C$ equals $A$.
  To match use of $W$$=$$L$ in the case of the RCW caches,
  we assume an implementation of
  LRU with \emph{Cache on $k^{th}$ request}
  in which, when the cache is full (as it is in steady state),
  $W$ is dynamically set to the number of requests since the
  ``least recently requested'' object currently
  in the cache was last requested.
  Similar to an RCW cache with $W$$=$$L$,
  this choice ensures that an object remains a ``caching candidate''
  as long as it is requested
  at least as recently as the least recently requested object in the cache.
  For the simulation results reported here and in subsequent sections,
  each simulation was
  run for six million requests, with the statistics for the initial two million
  requests removed from the measurements.}

\revthree{}{The following observations stand out.
  First, for all cases (including larger $k$),
  the exact RCW results closely match the LRU simulation results.
  This shows that the RCW analysis (presented here) can be used
  as an effective method to approximate the performance of an LRU cache.}

\revthree{}{Second, the $\mathcal{O}(1)$ approximations are relatively accurate,
  for both cases where caching is quite effective ($\alpha$$=$$1$) and largely ineffective ($\alpha$$=$$0.5$).
  For relative insertion rates (not shown), the $\mathcal{O}(1)$ diverge somewhat
    more, but errors remain within 10\% for
    all cases except for the cases of (i) small cache sizes ($A/N < 0.001$) when $k$$=$$4$ and $\alpha$$=$$0.5$, and
    (ii) large cache sizes ($A/N > 0.1$) when $k$$=$$4$ and $\alpha$$=$$0.5$.}
    \revfour{}{Here, it is important to note that accurate approximations for insertion rates with larger $k$ are a more difficult problem, since summations including larger powers of $(1 - p_i)$ need to be approximated (see equations (\ref{kW=L})). This is particularly an issue for $\alpha$$=$$0.5$ since the basic summation approximations we use (Section 3.2) for $\alpha = 0.5$ place tighter constraints on the powers of $(1 - p_i)$ for which our basic summation approximations can be expected to be most accurate (powers substantially less than $2N$ rather than substantially less than $N(\ln N + \gamma)$ as is the case for $\alpha$$=$$1$; \revDerek{e.g.,}{} compare
    \revDerek{approximation}{the assumptions used for} equations (\ref{appendixsqoneminuspLapprox}) and (\ref{appendixsqponeminuspLapprox}) versus equations (\ref{appendixoneminuspLapprox}) and (\ref{appendixponeminuspLapprox})).}
    \revthree{}{For $\alpha$$=$$1$, the relative insertion rate errors remain within 5\%, and for $k$$=$$1$ (regardless of $\alpha$) the errors are within 0.3\%.}
    \revfour{}{When discussing these $\mathcal{O}(1)$ approximations, it is also important to remember that the exact RCW results (which as per our first observation provide good results for all cases) can be used to approximate LRU.}

  \revthree{}{Third, the gap in hit rate between the policies and the upper bound
  is substantial with $k$$=$$1$ (regular LRU),
  narrows with $k$$=$$2$, and is almost eliminated with $k$$=$$4$,
  leaving little room for further hit rate improvements.}

\begin{figure}[t]
  \centering
  \subfigure[$\alpha=1$]{
    \includegraphics[trim = 0mm 6mm 0mm 0mm, width=0.24\textwidth]{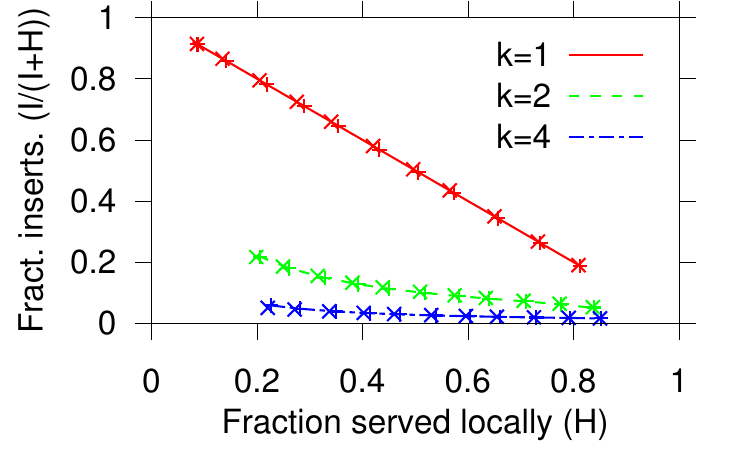}}
  \hspace{-14pt}
  \subfigure[$\alpha=0.5$]{
    \includegraphics[trim = 0mm 6mm 0mm 0mm, width=0.24\textwidth]{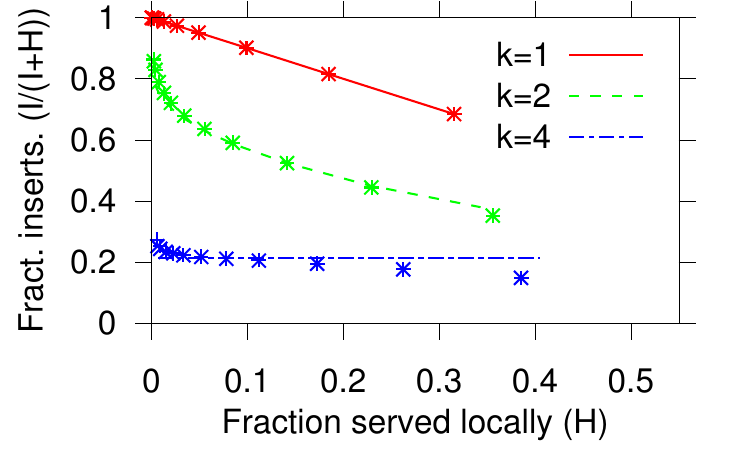}}
  \vspace{-12pt}
  \caption{Tradeoff curves for different {\em Cache on $k^{th}$ request} policies.  Here,
    $N=100,000$ and we use the following markers: ``RCW, exact'' (+), ``RCW, approx'' (line), and LRU ($\times$).}
  \label{fig:accuracy-tradeoff}
  \vspace{-12pt}
\end{figure}

\revthree{}{While comparing the different figures suggest that
  further increasing $k$ beyond $k$$=$$2$ yields only small additional
  improvements in the hit rates,
  it should be noted that larger improvements are seen in the insertion fraction
  and that these improvements continue
  (at least with respect to relative rather than absolute differences)
  as $k$ is increased.  To illustrate this,
  Figure~\ref{fig:accuracy-tradeoff} shows the tradeoff
  between hit rate (on x-axis) and insertion fraction (y-axis) for different $k$.
  Note that with selective cache insertion policies (i.e., larger $k$),
  the same hit rate can be achieved with a much lower insertion fraction.}


\section{Dynamic Instantiation Analysis}\label{transientperiodanalysis}

\revfour{}{In the following, we present an exact analysis for the cache hit rates and insertion rates during the transient periods for the cases of {\em Cache on k$^{th}$ request} when $k=1$ (Section 4.1) and $k \ge 2$ (Section 4.2), respectively, and derive the corresponding $\mathcal{O}(1)$ approximations (Section 4.3).  Performance results are presented in Section 4.4.} 

\subsection{Cache on 1st Request}

Consider now the case where the cache is allocated for only
a portion of the time period, and is initially empty when instantiated.
With \emph{Cache on $1^{st}$ request}, after the first $L$ requests
following instantiation,
the cache will have the occupancy probabilities derived earlier for
the ``always-on'' case in
\revthree{Section~\ref{cacheonfirst},}{Section~\ref{alwaysonanalysis},}
and so for requests following
the first $L$ requests
the analysis in
\revthree{Section~\ref{cacheonfirst}}{Section~\ref{alwaysonanalysis}}
can be used.
The average insertion rate over the first $L$ requests (the transient
period)
is given by the expression for the average number $A$ of objects
in cache
\revthree{from~(\ref{1exact}) in \revthree{Section~\ref{cacheonfirst},}{Section~\ref{alwaysonanalysis},}}{from~(\ref{1exact}),}
divided by $L$.
Denoting the average hit rate during the transient period by $\bar{H}_{\textrm{transient}}$, this gives:
{\footnotesize
\begin{equation}\label{1hitrtransient}
	\bar{H}_{\textrm{transient}} = 1 - \frac{A}{L} = 1 - \frac{N - \sum_{i=1}^N (1 - p_i)^L}{L},
\end{equation}}
\revthree{and from the equation for $H$ in~(\ref{1exact}),}{and from equation~(\ref{1exact}) for $H$,}
assuming that $\int_{t_a}^{t_d} \lambda (t) \textrm{d}t \geq L$,
{\footnotesize
\begin{equation}\label{1hitrinstantiation}
\bar{H}_{t_a : t_d} =
\frac{L \bar{H}_{\textrm{transient}} + \left( \int_{t_a}^{t_d} \lambda (t) \textrm{d}t - L \right) \left(1 - \sum_{i=1}^N p_i ( 1 - p_i)^L\right)  }{\int_{t_a}^{t_d} \lambda (t) \textrm{d}t}.
\end{equation}}
\revthree{}{Finally, for \emph{Cache on $1^{st}$ request},}
$\bar{I}_{\textrm{transient}}$ and $\bar{I}_{t_a : t_d}$ are given simply by
$1$$-$$\bar{H}_{\textrm{transient}}$ and $1$$-$$\bar{H}_{t_a : t_d}$.

\subsection{Cache on kth Request ($k \geq 2$)}\label{cacheonkthtransient}

As described in
\revthree{Section~\ref{cachepolicies},}{Section~\ref{systemdescriptionandmetrics},}
\emph{Cache on $k^{th}$ request} requires maintenance of
state information regarding
\revthree{``caching candidates".
Also, any RCW cache requires that the system maintain state
information regarding cached objects, so that such an object
can be evicted if it is not accessed over a window consisting of the most
recent $L$ requests.}{``caching candidates'' and all currently cached objects.}
We assume that when a cache using \emph{Cache on $k^{th}$ request}
is deallocated, the state information of both types
is transferred to the upstream
system to which requests will now be directed.
The upstream system maintains and updates this state information
when receiving requests that the cache would have received had it
been allocated,
and transfers it back when the cache is instantiated again.
Therefore, although the cache is initially empty when instantiated,
it can use the acquired state information to selectively cache newly requested
objects, caching a requested object not present in the cache,
whenever that object should be in (or be put in)
the cache according to its state
information.
Note that after the first $L$ requests
following instantiation,
the cache will have the cache occupancy probabilities derived earlier for
the ``always-on'' case,
and so for requests following
the first $L$ requests the analysis in
\revthree{Sections~\ref{cacheonsecond} and~\ref{cacheonkth}}{Section~\ref{alwaysonanalysis}}
can be used.

Note that over the transient period consisting
of the first $L$ requests,
no objects are removed from the cache.
The average insertion rate during the transient period
$\bar{I}_{\textrm{transient}}$ is
therefore given by the average number $A$ of objects in cache
(from~(\ref{2objs}) for $k$$=$$2$ and~(\ref{kobjs}) for general $k$),
divided by $L$.
Under the assumption that
$\int_{t_a}^{t_d} \lambda (t) \textrm{d}t \geq L$,
it is then straightforward to combine $\bar{I}_{\textrm{transient}}$
with the always-on insertion rate from Section~\ref{alwaysonanalysis}
to obtain $\bar{I}_{t_a : t_d}$.

The average hit rate during the transient period is given by one
minus the average transient period insertion rate,
minus the average probability that a requested object
is not present in the cache and should not be inserted.
Recall that the cache receives up-to-date state information when
instantiated,
\revthree{and caches a requested object not present in the cache whenever that
  object should be in (or be put in) the cache according to its state information.}{and a requested object is cached according
  to this state information.}
Therefore,
a requested object is not present in the cache and should not be
inserted, if and only if
it would not be in the cache and would not be inserted into the cache
on this request with an always-on cache.
The probability of this case is equal to one minus the hit rate
for an always-on cache minus the insertion rate for an always-on cache.
The above implies that
the average hit rate during the transient period,
$\bar{H}_{\textrm{transient}}$,
is given by
the always-on cache hit rate (in~(\ref{khit})) plus
the always-on cache insertion rate (in~(\ref{kins}))
minus the average transient period insertion rate
\revthree{$\bar{I}_{\textrm{transient}}$.}{$\bar{I}_{\textrm{transient}}$;
  i.e., $\bar{H}_{\textrm{transient}} = H + I - A/L$.}
Under the assumption that
$\int_{t_a}^{t_d} \lambda (t) \textrm{d}t \geq L$,
it is then straightforward to combine $\bar{H}_{\textrm{transient}}$
with the always-on hit rate
\revthree{from Section~\ref{alwaysonanalysis}}{(in (\ref{khit}))}
to obtain $\bar{H}_{t_a : t_d}$.


\subsection{Dynamic Instantiation Approximations}\label{transientperiodanalysis-appendix}

\revExt{}{We next derive and analyze our
easy-to-compute $\mathcal{O}(1)$-approximations
for the transient period.  A summary of these results
are provided in Table~\ref{tab:summary-approximations}.}

\subsubsection{Cache on 1st Request}

\revDerek{The ratio of the average cache hit rate over the transient period to the
hit rate once the cache has filled can yield substantial
insight into the impact of
the transient period on performance.}{}
For the special case of a Zipf object popularity distribution with
$\alpha = 1$, using~(\ref{1objs}) to substitute for $A$ in
$\bar{H}_{\textrm{transient}} = 1 - \frac{A}{L}$
yields
{\footnotesize
\begin{equation}
  \bar{H}_{\textrm{transient}} \approx \frac{\ln ( L/(\ln N + \gamma)) + 2\gamma - 1 - L/(2N(\ln N + \gamma))}{\ln N + \gamma}.
\end{equation}}
\revDerek{}{The ratio of the average cache hit rate over the transient period to the
hit rate once the cache has filled can yield substantial
insight into the impact of
the transient period on performance.}
\revDerek{The ratio of the average cache hit rate over the transient period
to the hit rate once the cache has filled (given by expression~(\ref{1hit})) is therefore approximately}
{In this case, from the above expression and expression~(\ref{1hit}) this ratio is approximately}
$1 - (1 - L/(2N(\ln N + \gamma))) / (\ln N + \gamma)$.

For $\alpha = 0.5$, using~(\ref{sq1objs}) to substitute for $A$
yields
{\footnotesize
\begin{equation}
  \bar{H}_{\textrm{transient}} \approx \frac{L}{4N} \left( \ln ((2N)/L) + \frac{L}{6N} + \frac{3}{2} - \gamma \right).
\end{equation}}
In this case the ratio of the average hit rate
over the transient period to the hit rate once the cache has filled
(given in~(\ref{sq1hit})) is between 0.5 and 0.7
(for $0 < L < 2N$), substantially smaller than for
$\alpha = 1$.

\begin{figure*}[t]
  \centering
  \subfigure[$k=1, \alpha=1$]{
    \includegraphics[trim = 1mm 4mm 4mm 0mm,  width=0.29\textwidth]{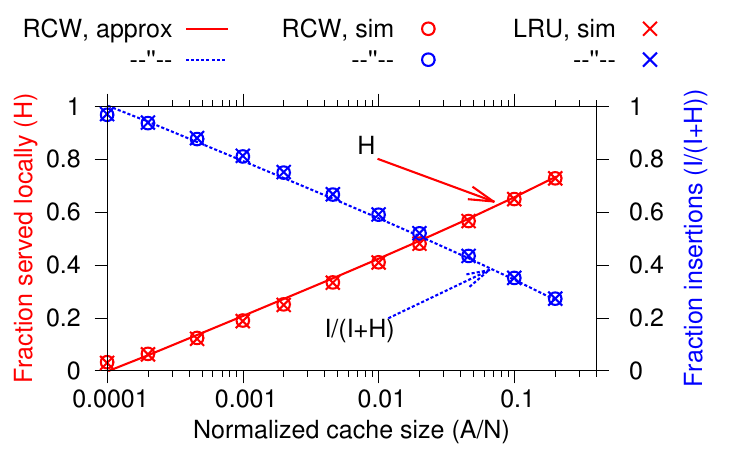}}
  \subfigure[$k=2, \alpha=1$]{
    \includegraphics[trim = 1mm 4mm 4mm 0mm,  width=0.29\textwidth]{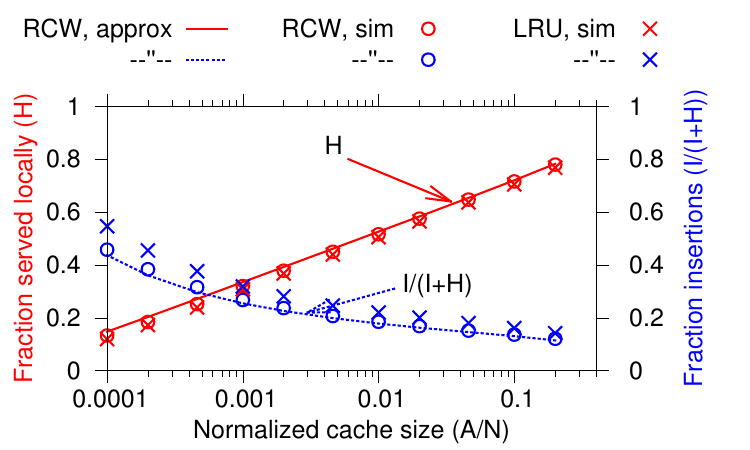}}
  \subfigure[$k=4, \alpha=1$]{
    \includegraphics[trim = 1mm 4mm 4mm 0mm,  width=0.29\textwidth]{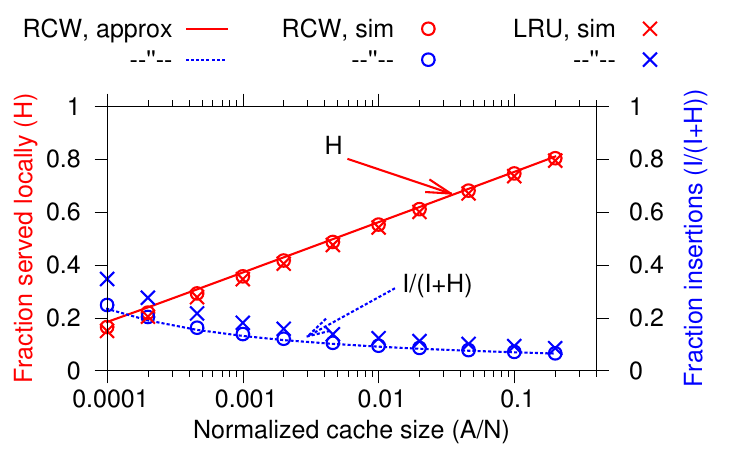}}
  \subfigure[\vspace{-14pt}$k=1, \alpha=0.5$]{
    \includegraphics[trim = 1mm 2.2mm 4mm 16mm, clip, width=0.29\textwidth]{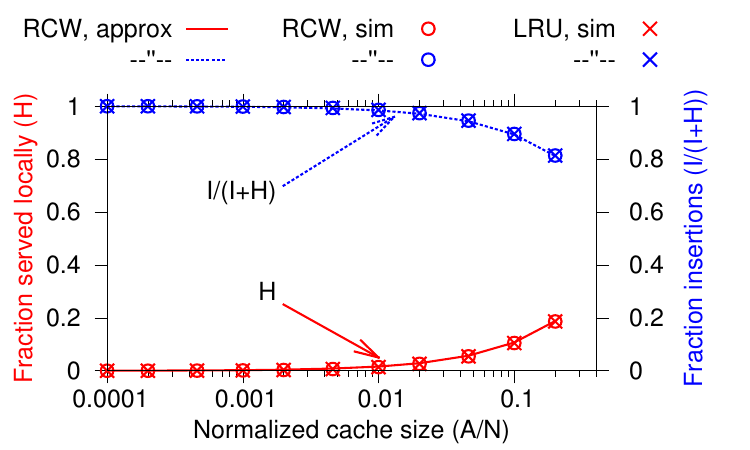}}
  \subfigure[$k=2, \alpha=0.5$]{
    \includegraphics[trim = 1mm 2.2mm 4mm 16mm, clip, width=0.29\textwidth]{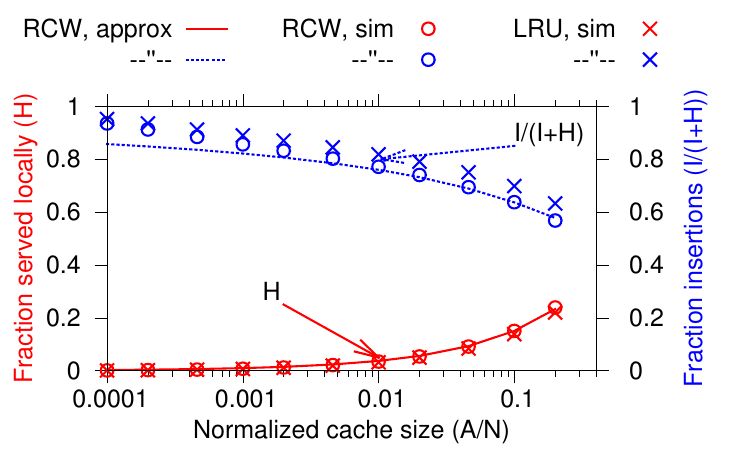}}
  \subfigure[$k=4, \alpha=0.5$]{
    \includegraphics[trim = 1mm 2.2mm 4mm 16mm, clip, width=0.29\textwidth]{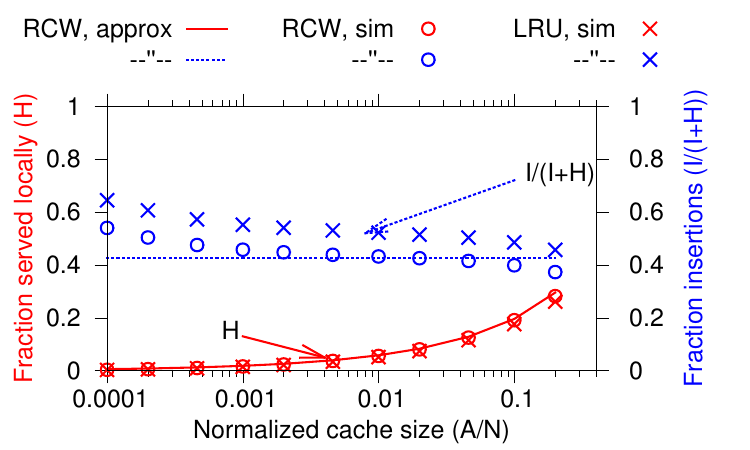}}
  \vspace{-12pt}
  \caption{\revtwo{}{Transient period results.  Performance of {\em Cache on $k^{\textrm{th}}$ request} ($N$ = 100,000)
      during transient period.}}
  \label{fig:transient}
  \vspace{-14pt}
\end{figure*}

\subsubsection{Cache on kth Request ($k \geq 2$)}\label{cacheonkthtransient-appendix}

\revDerek{Consider now
\revrev{}{approximations for}
the case of $W = L$ and
a Zipf object popularity distribution.
For $\alpha = 1$ and $k=2$,}{For $k=2$, $W=L$, and a Zipf object popularity distribution with $\alpha = 1$,}
applying the expressions for $H$ and $I$ in~(\ref{2approx-H}) and~(\ref{2approx-I}), and using
the expression for $A$ in~(\ref{2approx-A}) to substitute for $A$ in the transient period insertion rate $A/L$,
yields an approximation for the average hit rate during the transient period of
{\footnotesize
  \begin{align}
    \bar{H}_{\textrm{transient}} &\approx \frac{\ln (L / (\ln N + \gamma)) + 2 \gamma - \ln 2 }{\ln N + \gamma} \nonumber \\
    & ~~~~~~~ + \frac{\ln 2 - L/(N (\ln N + \gamma))}{\ln N + \gamma} - \frac{2 \ln 2 - L/(N(\ln N + \gamma))}{\ln N + \gamma} \nonumber \\
    &=  \frac{\ln (L/(\ln N + \gamma)) + 2\gamma - 2\ln 2}{\ln N + \gamma}.
\end{align}}
The ratio of the cache hit rate over the transient period to the
hit rate once the cache has filled (given in~(\ref{2approx-H}))
is therefore approximately $1 - (\ln 2)/(\ln (L/(\ln N + \gamma )) + 2 \gamma - \ln 2)$.
In contrast, for $\alpha = 0.5$ and $k=2$,
applying the expressions for the hit and insertion rates in~(\ref{sq2-H}) and~(\ref{sq2-I}),
and the expression for $A$ in~(\ref{sq2-A}) to substitute for $A$ in the
transient period insertion rate $A/L$,
yields an approximation for the average hit rate over the
transient period of
{\footnotesize
\begin{equation}
  \bar{H}_{\textrm{transient}} \approx \frac{L}{N} \left( \ln 2 - \frac{L}{8N} - \frac{1}{4} \right).
\end{equation}}
From comparison with the hit rate expression in~(\ref{sq2-H}),
the ratio of the average cache hit rate over the
transient period to the hit rate once the cache has filled is
between about 0.64 and 0.72
(considering here $0 < L < N$),
substantially smaller than for $\alpha = 1$.
Results similar in nature are obtained for $k \geq 3$,
applying~(\ref{khitapprox}),~(\ref{kinsapprox}), and~(\ref{kobjsapprox})
for $\alpha = 1$, and~(\ref{sqkhitr}),~(\ref{sqkins}), and~(\ref{sqkobjs}) for $\alpha = 0.5$.

\subsection{Transient Period Performance Results }

\revtwo{}{Figure~\ref{fig:transient} shows sample results for
the transient period when using \emph{Cache on $k^{th}$ request} with
  different $k=1,2,4$ and Zipf with $\alpha = 0.5$ or 1.  In all experiments,
  we used $W$$=$$L$ and show results only for the transient period itself.
  For the analytic expressions,
  we used the $\mathcal{O}(1)$ approximations
  \revExt{\revthree{from the prior sub-sections.}{for each metric (see appendix).}}{for each metric.}
  For the simulations,
  we start with an empty cache,
  and simulate the system until the system reaches steady-state conditions.
  At that time, we empty the cache and begin a new transient period.
  This is repeated for 2,000 transient periods or until we have simulated 6,000,000 requests,
  whichever occurs first,
  and statistics are reported based on fully completed transient periods.
  With these settings,
  each data point was calculated based on at least 17 transient periods.
  (This occurred with $A/N$$=$$0.2$, $k$$=$$4$, and $\alpha$$=$$1$.)
  To improve readability, as in prior figures,
  confidence intervals are not included.  However,
  in general, the confidence intervals are tight
  (e.g., $\pm 0.0016$ for the data point mentioned above).}

\revthree{\revtwo{}{For the RCW simulations,
  similar to the modeling assumptions,
  the system maintains state about requests to each object at all times.
  In particular,
  while we start each transient period with an empty cache,
  we keep the (prior) information about the number of consecutive times that each object have
  been requested within $W$ of a prior request to the same object.}}{For the RCW simulations,
  the system maintains the same state information and operates in the same way as described
  for our analysis assumptions.}
\revfour{As in the steady-state simulations, for the corresponding LRU cache, the capacity $C$ was set equal to $A$, and when the cache is full, $W$ is dynamically set to the number of requests since the ``least recently requested'' object currently in the cache (and with at least $k$ requests within $W$ of each other) was last requested. To reach steady state conditions, the cache must be filled completely with objects
  requested at least $k$ times during that period.}{As in the steady-state simulations, for the corresponding LRU cache, the capacity $C$ was set equal to $A$.  However, rather than dynamically setting $W$ to the number of requests since the ``least recently requested” object currently in the cache was last requested (corresponding to $W=L$), as was done for the steady-state simulations, we dynamically inflate $W$ by one for each request being made during the transient period. This ensures that no per-object counters are reset during the transient phase, regardless of how long the transient period is. The cache reaches steady state conditions when it is filled completely.  Otherwise the implementation is the same as for the RCW simulations, including that objects are added to the cache whenever their counter reaches $k$.}

\revtwo{}{The transient results very much resemble
  the steady-state results.
  For example, the tradeoff curves in Figure~\ref{fig:transient} are very similar
  to those observed in
  \revthree{Figures~\ref{fig:accuracy-k1}-\ref{fig:accuracy-k4},}{Figure~\ref{fig:accuracy-k124},}
  and the analytic approximations again nicely match the simulated RCW values for most
  \revthree{instances.}{instances (which themselves nicely match the exact analysis results).}
  Most importantly, there is a very good match for all hit rate results
  (red curves/markers: RCW approximations, RCW simulations, and LRU simulations);
  the metric that we will use in the optimization models (Section~\ref{optimizationmodel})
  and the evaluation thereof (Section~\ref{comparisons}).
  Substantive differences between the RCW simulations and analytic approximations
  are observed only
  for the insertion fraction metric
  (shown in blue)
  when using very small cache sizes
  (e.g., $A/N$ less than 0.001) when $\alpha$$=$$0.5$.
  When $k$$=$$4$ and $\alpha$$=$$0.5$, we also observe some noticeable differences in the insertion fraction
  between RCW and LRU.  This may suggest that when $k$ is large,
  RCW is a worse approximation for LRU (as we compare them)
  during transient periods than during steady state.
\revfour{}{The main reason for this is related to the difficulty in selecting a good $W$ for the LRU cache during the transient period. By ensuring that all requests made during the transient period increase a per-object request counter,  our implementation of the transient period operation of the LRU cache provides a clear implementation of a {\em Cache on k$^{th}$ request} policy. However, with $k>1$, this implementation ends up adding objects to the cache at a higher rate than the RCW cache.  This results in larger transient-period insertion fractions than with RCW when $k > 1$, with the difference becoming larger for larger $k$.
A less aggressive policy may be to freeze $W$ during the transient period (based on the last observed value before the cache was released) and only use our current policy for the initial transient period (when $W$ is still unknown). However, this approach can become sensitive to the value that $W$ has at the moment when a cache is released.  Cache performance comparisons of this and other alternative LRU policy variations provide interesting future work but are outside the scope of this paper.}
  \revfour{Yet,}{Yet, for the purpose of this paper,} 
  for all considered $k$ and $\alpha$,
  we find the approximations sufficiently accurate to justify using them
  for our optimization of dynamic cache instantiation.
  Again, in the following sections, we will leverage the (more accurate) hit rate results.}

\section{Optimization Models for Dynamic Instantiation}\label{optimizationmodel}

Consider now the problem of jointly optimizing the capacity $C$
of a dynamically instantiated cache,
and the interval over which the cache is allocated, so as to minimize
the cache cost subject to achieving a target fraction of requests
$H_{\min}$ ($0 < H_{\min} < 1$) that will be served locally:
{\footnotesize
\begin{equation}\label{optfun}
	\textrm{minimize} \quad (t_d - t_a) (C + b),
\end{equation}}
{\footnotesize
\begin{equation}
	\textrm{subject to} \quad \frac{\bar{H}_{t_a : t_d} \int_{t_a}^{t_d} \lambda (t) \textrm{d}t}{\int_0^T \lambda (t) \textrm{d}t} \geq H_{\min}. \nonumber
\end{equation}}
Note that a smaller cache has the advantages of a shorter transient
period until it fills and lower cost per unit time, while a larger cache has
the advantage of a higher hit rate once filled.
\revthree{It is assumed for convenience in the following}{For convenience, in the following we assume}
that
\revthree{$\lambda(t) > 0$ for all $t$.}{$\lambda(t) > 0, \forall t$.}

\revfour{}{In this section, using the above optimization formulation, we derive \revDerek{lower bounds}{a cost lower bound} (Section 5.1),
\revDerek{present approximation-based bound}{present bounds specialized to Zipf popularity distributions} (Section 5.2), and present both policy-based cost optimizations (Section 5.3) and their corresponding approximate cost optimizations (Section 5.4).  These results are then used in the performance evaluation of different dynamic instantiation solutions presented in Section 6.}

\subsection{Lower Bound}\label{lowerboundderivation}

A lower bound on cost can be obtained by using an upper bound for
the average hit rate over the cache allocation interval.
One such bound can be obtained by assuming that there is a
hit whenever the requested object is one that has been requested
previously, since the cache was allocated.
We apply this bound to obtain a lower bound on the duration of the
cache allocation interval.
Another bound is the hit rate when the $\lfloor{C}\rfloor$ most
popular objects are present in the cache.
We apply this bound to the more constrained optimization problem
that results from our use of the first bound.

Denote by $\bar{H}_R$ the average hit rate over
the first $R$ requests after the cache has been allocated.
At best, request $r$, $1 \leq r \leq R$ is a hit if and only
if the requested object was the object requested by one or more of
the $r-1$ earlier requests, giving:
{\footnotesize
\begin{equation}\label{hitrbound}
\bar{H}_R \leq \frac{1}{R} \left( \sum_{r=1}^{R} \sum_{i=1}^N p_i (1 \scalebox{0.75}[1.0]{\( - \)} (1 \scalebox{0.75}[1.0]{\( - \)} p_i)^{r \scalebox{0.75}[1.0]{\( - \)} 1}) \right)
= 1 - \frac{1}{R} \left( N  -   \sum_{i=1}^N (1 \scalebox{0.75}[1.0]{\( - \)} p_i)^R \right).
\end{equation}}
Since this is a concave function of $R$, we can bound the average hit
rate over the cache allocation interval
by setting $R = \int_{t_a}^{t_d} \lambda (t) \textrm{d}t$,
the expected value of the number of requests within this
interval.
Applying this bound to the hit rate constraint in~(\ref{optfun}) yields
{\footnotesize
\begin{equation}
\left( \frac{\int_{t_a}^{t_d} \lambda (t) \textrm{d}t}{\int_0^T \lambda (t) \textrm{d}t} \right)
\left( 1 - \frac{1}{R} \left( N - \sum_{i=1}^N (1-p_i)^R \right) \right)
\geq H_{\min},
\end{equation}}
implying that
{\footnotesize
\begin{equation}\label{lowerboundtd-ta}
\int_{t_a}^{t_d} \lambda (t) \textrm{d}t + \sum_{i=1}^N (1-p_i)^{\int_{t_a}^{t_d} \lambda (t) \textrm{d}t} \geq
\left( \int_0^T \lambda (t) \textrm{d}t \right) H_{\min} + N.
\end{equation}}
Given that we choose $t_a$ and $t_d$ as the beginning and end, respectively,
of a time interval
\revthree{of duration $t_d - t_a$ with}{with}
the largest
average request rate,
the left-hand side is a strictly increasing function of
$t_d - t_a$, as can be verified by taking the derivative with respect
to $\int_{t_a}^{t_d} \lambda (t) \textrm{d}t$,
noting that this derivative is minimized for minimum
$\int_{t_a}^{t_d} \lambda (t) \textrm{d}t$ (which is at least one,
in the region of interest), and using the fact
that $- \ln (x)$ is a convex function.
Therefore, for any particular workload this relation defines a lower bound
$D_l$ for the interval duration $t_d - t_a$.

Applying now the upper bound on hit rate from
when the $\lfloor{C}\rfloor$ most
popular objects are present in the cache, gives
the following optimization problem:
{\footnotesize
\begin{equation}\label{lowerbound}
\textrm{minimize} \quad (t_d - t_a) (C + b),
\end{equation}
\begin{equation}
\textrm{subject to} \quad \left( \frac{\int_{t_a}^{t_d} \lambda (t) \textrm{d}t}{\int_0^T \lambda (t) \textrm{d}t} \right) \sum_{i=1}^{\lfloor{C}\rfloor} p_i \geq H_{\min}, \quad D_l \leq t_d- t_a \leq T. \nonumber
\end{equation}}
Solution of this
\revthree{optimization problem}{problem}
yields a lower bound on cost.
\revExt{}{Specializations of this problem for the cases of Zipf popularity distributions with
  $\alpha$ = 1 and 0.5 are developed next.}


\subsection{\revDerek{Approximation-based Bounds}{Bounds for Zipf Popularity Distributions}}\label{sec:bound-optimization}

\subsubsection{Zipf with {\large $\alpha$} = 1}

Consider now the special case of a Zipf object popularity distribution with
parameter $\alpha = 1$, and denote the normalization constant
$\sum_{i=1}^N 1/i$ by $\Omega$.
In the case that $R < (N+1) (\ln (N+1) + \gamma)$, we have
{\footnotesize
  \begin{align}
    & \sum_{i=1}^N (1-p_i)^R = \sum_{i=1}^N \left(1 - \frac{1}{i\Omega} \right)^{(i\Omega) \frac{R}{i\Omega}} \leq \sum_{i=1}^N e^{-R/(i\Omega)} \nonumber \\
    & ~~~~ < \int_1^{N+1} e^{-R/(x\Omega)} \textrm{d}x
    < \int_1^{N+1} e^{-R/(x(\ln (N+1)+\gamma))} \textrm{d}x \nonumber \\
    & ~~~~ < (N+1) - R + \frac{R}{\ln (N+1) + \gamma} \left(\ln \left( \frac{R}{\ln (N+1) + \gamma} \right) + 2\gamma - 1\right),
\end{align}}
where the second last inequality uses $\Omega < \ln (N+1) + \gamma$,
and the last inequality follows from the Taylor series
expansion (as in~(\ref{taylor}) in the Section~\ref{sec:summation-approx})
under the assumption that $R < (N+1) (\ln (N+1) + \gamma)$.
Using $R = \int_{t_a}^{t_d} \lambda (t) \textrm{d}t$,
under the assumption that $R = \int_{t_a}^{t_d} \lambda (t) \textrm{d}t < (N+1) (\ln (N+1) + \gamma)$,
we can substitute into~(\ref{lowerboundtd-ta}) to obtain:
{\footnotesize
  \begin{equation}
    \frac{\int_{t_a}^{t_d} \lambda (t) \textrm{d}t}{\ln (N+1) + \gamma} \left( \ln \left( \frac{\int_{t_a}^{t_d} \lambda (t) \textrm{d}t}{\ln (N+1) + \gamma} \right) + 2\gamma - 1\right)
    \geq \left( \int_0^T \lambda (t) \textrm{d}t \right) H_{\min} - 1.
\end{equation}}
When the right-hand side of this relation is positive,
which it is for parameters of interest,
the left-hand side must be a strictly increasing function of $t_d - t_a$.
Under
\revrev{our}{the}
assumption that
$\int_{t_a}^{t_d} \lambda (t) \textrm{d}t < (N+1) (\ln (N+1) + \gamma)$,
a lower bound $D_l^\prime$ for $t_d - t_a$
can therefore be obtained from
this relation for any particular workload of interest
(setting $D_l^\prime = \infty$ if no value for
$t_d - t_a \leq T$ satisfies this relation).
Denoting by $D_l^{\prime \prime}$ the maximum value of $t_d - t_a$
such that $\int_{t_a}^{t_d} \lambda (t) \textrm{d}t \leq
(N+1) (\ln (N+1) + \gamma)$,
with $D_l^{\prime \prime} = \infty$ if this still holds for
$t_d - t_a = T$,
a lower bound $D_l$ for $t_d - t_a$ is given by:
{\footnotesize
\begin{equation}\label{Dl}
  D_l = \min \left[ D_l^\prime, D_l^{\prime \prime} \right].
\end{equation}}

Also, for the special case of a Zipf object popularity distribution with
parameter $\alpha = 1$ and $C \leq N$,
$\sum_{i=1}^{\lfloor{C}\rfloor} p_i < (\ln (C+1) + \gamma)/(\ln N + \gamma)$.
Applying this bound to the hit rate
constraint in~(\ref{lowerbound}) yields the following optimization problem:
{\footnotesize
\begin{equation}\label{lowerbound2}
  \textrm{minimize} \quad (t_d - t_a) (C + b),
\end{equation}}
subject to
{\footnotesize
\begin{align}
  \left( \frac{\int_{t_a}^{t_d} \lambda (t) \textrm{d}t}{\int_0^T \lambda (t) \textrm{d}t} \right)
  \frac{\ln (C+1) + \gamma}{\ln N + \gamma} \geq H_{\min}, \nonumber \\
  C \leq N,  \quad D_l \leq t_d - t_a \leq T, \nonumber
\end{align}}
where $D_l$ is given by~(\ref{Dl}).
This optimization problem can be further specialized to any particular
workload of interest by specifying,
as a function of the duration $D = t_d - t_a \leq T$,
the average request rate that the cache would experience should
it be allocated for the interval of duration $D$, within the time
period under consideration, with
the highest average request rate.
It is then straightforward to solve the optimization problem to
any desired degree of precision.
The computational cost of evaluating the optimization function
and checking the constraints is $\mathcal{O}(1)$,
and it is feasible to simply search over all choices of
$C$ and the duration $t_d - t_a$ of the cache allocation interval,
at some desired granularity,
to find the choices that satisfy the constraints (should any
such choices exist) with lowest cost.

\subsubsection{Zipf with {\large $\alpha$} = 0.5}

For a Zipf object popularity distribution with $\alpha = 0.5$,
the normalization constant $\Omega = \sum_{i=1}^N 1/\sqrt{i}$.
In the case that $R < 2 (N+1)$ we have
{\footnotesize
  \begin{align}
    & \sum_{i=1}^N (1-p_i)^R = \sum_{i=1}^N \left(1 - \frac{1}{(\sqrt{i}) \Omega} \right)^{((\sqrt{i})\Omega) \frac{R}{(\sqrt{i})\Omega}} \nonumber \\
    & ~~~ \leq \sum_{i=1}^N e^{\frac{-R}{(\sqrt{i})\Omega}}
    < \int_1^{N+1} e^{-R/((\sqrt{x})\Omega)} \textrm{d}x
    < \int_1^{N+1} e^{-R/(2 \sqrt{x} \sqrt{N+1})} \textrm{d}x \nonumber \\
    & ~~~ < N+1 - R + \frac{R^2}{4(N+1)} \left( \ln \left(\frac{2(N+1)}{R} \right) + \frac{R}{6(N+1)} + \frac{3}{2} - \gamma \right),
\end{align}}
where the second last inequality uses $\Omega < 2 \sqrt{N+1}$,
and the last inequality follows from the Taylor series
expansion (as in~(\ref{sqtaylor}) in the Section~\ref{sec:summation-approx})
under the assumption that $R < 2 (N+1)$.
Using $R = \int_{t_a}^{t_d} \lambda (t) \textrm{d}t$,
under the assumption that $R = \int_{t_a}^{t_d} \lambda (t) \textrm{d}t < 2 (N+1)$,
we can substitute into~(\ref{lowerboundtd-ta}) to obtain:
{\footnotesize
  \begin{align}
    & \frac{\left(\int_{t_a}^{t_d} \lambda (t) \textrm{d}t \right)^2}{4(N+1)} \left( \ln \left(\frac{2(N+1)}{\int_{t_a}^{t_d} \lambda (t) \textrm{d}t} \right) + \frac{\int_{t_a}^{t_d} \lambda (t) \textrm{d}t}{6(N+1)} + \frac{3}{2} - \gamma \right) \nonumber \\
    & ~~~~~~~~~~~~~~~~~~~~~~~~~~~~~~~~~~~~~~~\geq \left( \int_0^T \lambda (t) \textrm{d}t \right) H_{\min} - 1.
\end{align}}
The left-hand side of this relation is a strictly increasing function
of $t_d - t_a$.
Under
\revrev{our}{the}
assumption that
$\int_{t_a}^{t_d} \lambda (t) \textrm{d}t < 2 (N+1)$,
a lower bound $D_l^\prime$ for $t_d - t_a$
can therefore be obtained from
this relation for any particular workload of interest
(setting $D_l^\prime = \infty$ if no value for
$t_d - t_a \leq T$ satisfies this relation).
Denoting by $D_l^{\prime \prime}$ the maximum value of $t_d - t_a$
such that $\int_{t_a}^{t_d} \lambda (t) \textrm{d}t \leq
2 (N+1)$,
with $D_l^{\prime \prime} = \infty$ if this still holds for
$t_d - t_a = T$,
a lower bound $D_l$ for $t_d - t_a$ is given by
$D_l = \min \left[ D_l^\prime, D_l^{\prime \prime} \right]$.

Also, for the special case of a Zipf object popularity distribution with
parameter $\alpha = 0.5$ and $C \leq N$,
{\footnotesize
\begin{equation}\label{sqLFU}
  \sum_{i=1}^{\lfloor{C}\rfloor} p_i < \frac{2 \sqrt{C + \frac{1}{2}} - \sqrt{2}}{2 \sqrt{N+1} - 2} = \frac{\sqrt{C + \frac{1}{2}} - \frac{1}{\sqrt{2}}}{\sqrt{N+1} - 1}, \quad C \leq N.
\end{equation}}
Applying the bound in~(\ref{sqLFU}) to the hit rate
constraint in~(\ref{lowerbound}) yields the following optimization problem:
{\footnotesize
\begin{equation}\label{sqlowerbound2}
  \textrm{minimize} \quad (t_d - t_a) (C + b),
\end{equation}}
subject to
{\footnotesize
\begin{align}
  \left( \frac{\int_{t_a}^{t_d} \lambda (t) \textrm{d}t}{\int_0^T \lambda (t) \textrm{d}t} \right)
  \frac{\sqrt{C + \frac{1}{2}} - \frac{1}{\sqrt{2}}}{\sqrt{N+1} - 1} \geq H_{\min}, \nonumber \\
  C \leq N, \quad D_l \leq t_d - t_a \leq T. \nonumber
\end{align}}
As before,
it is straightforward to specialize this optimization problem to
any particular workload of interest, and to then solve it to any
desired degree of precision.


%
\subsection{\revthree{Cache on 1st Request}{Policy-based Cost Optimizations}}
\revthree{}{{\bf Cache on $1^{st}$ request:}}
For an LRU cache using
\revthree{\emph{Cache on $1^{st}$ request},}{this policy,}
equating the cache capacity $C$ to the average occupancy $A$ of
an RCW cache and applying~(\ref{1hitrinstantiation})
\revthree{yields the following optimization problem:}{yields:}
{\footnotesize
\begin{equation}
\textrm{minimize} \quad (t_d - t_a) (C + b),
\end{equation}}
{\footnotesize
\begin{equation}
\textrm{subject to} \quad
 C = N - \sum_{i=1}^N (1 - p_i)^L, \quad
L \leq \int_{t_a}^{t_d} \lambda (t) \textrm{d}t, \nonumber
\end{equation}}
{\footnotesize
      \begin{equation}
 ~~~~\frac{L - C + \left(\int_{t_a}^{t_d} \lambda (t) \textrm{d}t - L \right) \left(1 - \sum_{i=1}^N p_i ( 1 - p_i)^L \right)}{\int_0^T \lambda (t) \textrm{d}t } \geq H_{\min}. \nonumber
\end{equation}}
    \revExt{}{Similarly as for the lower bound, specializations of this problem for the cases of
      Zipf popularity distributions with $\alpha$ = 1 and 0.5 are developed in Section~\ref{sec:approx-optimization}.}

\revthree{\subsection{Cache on kth Request}}{{\bf Cache on $k^{th}$ request:}}
\revthree{For an LRU cache using \emph{Cache on $k^{th}$ request},}{Similarly,}
equating the
\revthree{cache capacity}{capacity}
$C$ to the average occupancy $A$ of
an RCW cache and applying the Section~\ref{cacheonkthtransient} analysis
yields the optimization problem:
{\footnotesize
\begin{equation}
\textrm{minimize}~~~~ (t_d - t_a) (C + b),
\end{equation}}
    subject to
{\footnotesize
\begin{equation}
    C = \sum_{i=1}^N \frac{1 - (1 - p_i)^L}{1 - (1 - p_i)^L + \frac{(1 - p_i)^L (1 - (1 - (1 - p_i)^W)^k)}{(1 - p_i)^W (1 - (1 - p_i)^W)^{k-1}}}, ~
    L \leq \int_{t_a}^{t_d} \lambda (t) \textrm{d}t, \nonumber
\end{equation}}
\vspace{-12pt}
{\footnotesize
\begin{align}
& \frac{\int_{t_a}^{t_d} \lambda (t) \textrm{d}t }{\int_0^T \lambda (t) \textrm{d}t } \sum_{i=1}^N p_i \frac{1 - (1 - p_i)^L}{1 - (1 - p_i)^L + \frac{(1 - p_i)^L (1 - (1 - (1 - p_i)^W)^k)}{(1 - p_i)^W (1 - (1 - p_i)^W)^{k-1}}} \nonumber \\
& ~~~~~ + \frac{L \sum_{i=1}^N p_i \frac{(1 - p_i)^L}{1 - (1 - p_i)^L + \frac{(1 - p_i)^L (1 - (1 - (1 - p_i)^W)^k)}{(1 - p_i)^W (1 - (1 - p_i)^W)^{k-1}}} - C }{\int_0^T \lambda (t) \textrm{d}t } \geq H_{\min}. \nonumber
\end{align}}
\revExt{}{Section~\ref{sec:approx-optimization} develops specializations of this problem for $W=L$ and the cases of Zipf popularity
  distributions with $\alpha$ = 1 and 0.5.}

\subsection{Approximate Cost Optimization}\label{sec:approx-optimization}

{\bf Cache on 1st Request:}
For the special case of a Zipf object popularity distribution with
parameter $\alpha = 1$,
applying~(\ref{1hit}) and~(\ref{1L}) this becomes:
{\footnotesize
\begin{equation}\label{1opt}
  \textrm{minimize} \quad (t_d - t_a) (C + b),
\end{equation}}
subject to
{\footnotesize
  \begin{align}
    & L = \frac{N^\beta (\ln N + \gamma)((1-\beta) \ln N + \gamma - 1)} {((1- \beta) \ln N + \gamma)((1-\beta) \ln N - \gamma + \ln ((1-\beta) \ln N + \gamma))}, \nonumber \\
    & ~~~~~~~ \beta = \frac{\ln C}{\ln N}, \quad ~~~~~~~~~~~~ L \leq \int_{t_a}^{t_d} \lambda (t) \textrm{d}t, \nonumber
\end{align}}
\vspace{-10pt}
       {\footnotesize
         \begin{align}
           & \frac{L-C+\left(\int_{t_a}^{t_d} \lambda (t) \textrm{d}t - L \right) \left(\frac{\ln (L/(\ln N + \gamma)) + 2\gamma - L/(N(\ln N + \gamma))}{\ln N + \gamma} \right)}{\int_0^T \lambda (t) \textrm{d}t } \geq H_{\min}.
           \nonumber
       \end{align}}
       Similarly,
       applying~(\ref{sq1hit}) and~(\ref{sq1L})
       yields the corresponding optimization problem for $\alpha = 0.5$.
       As before, it is feasible to simply search over all choices of
       $C$ and the duration $t_d - t_a$ of the cache allocation interval,
       at some desired granularity,
       to find the choices that satisfy the constraints (should any
       such choices exist) with lowest cost.

       {\bf Cache on kth Request:}
       For the special case of $W = L$ and a Zipf object popularity distribution with
       parameter $\alpha = 1$,
       applying the expressions for $H$ and $I$ in~(\ref{2approx-H}) and~(\ref{2approx-I}), and~(\ref{2L}),
       yields the following optimization problem for \emph{Cache on $2^{nd}$ request}:
 {\footnotesize
 \begin{equation}\label{2opt}
         \textrm{minimize}~~~~ (t_d - t_a) (C + b),
       \end{equation}}
       subject to
{\footnotesize
\begin{align}
         & L = \frac{C (\ln N + \gamma) (1+ C/(4(\ln 2)^2N) )}{2\ln 2}, \quad
         L \leq \int_{t_a}^{t_d} \lambda (t) \textrm{d}t, \nonumber \\
         & \frac{L \left( \frac{\ln 2 - L/(N (\ln N + \gamma))}{\ln N + \gamma} \right) - C}{\int_0^T \lambda (t) \textrm{d}t } \nonumber \\
         & ~~~~~~~ + \frac{\int_{t_a}^{t_d} \lambda (t) \textrm{d}t }{\int_0^T \lambda (t) \textrm{d}t } \left( \frac{\ln (L / (\ln N + \gamma)) + 2 \gamma - \ln 2 }{\ln N + \gamma} \right) \geq H_{\min}. \nonumber
       \end{align}}
       Similarly, applying~(\ref{khitapprox}),~(\ref{kinsapprox}), and~(\ref{kL})
       yields the corresponding optimization problem for
       $k \geq 3$, while applying the expressions for $H$ and $I$
       in~(\ref{sq2-H}) and~(\ref{sq2-I}), and~(\ref{sq2L}) ($k=2$),
       and ~(\ref{sqkhitr}),~(\ref{sqkins}),~(\ref{sqkL}), and~(\ref{sq3L}) ($k \geq 3$)
       yield the corresponding optimization problems for $\alpha = 0.5$.

       \begin{figure*}[t]
         \centering
         \subfigure[Varying $b$]{
           \includegraphics[trim = 1mm 10mm 4mm 0mm, width=0.30\textwidth]{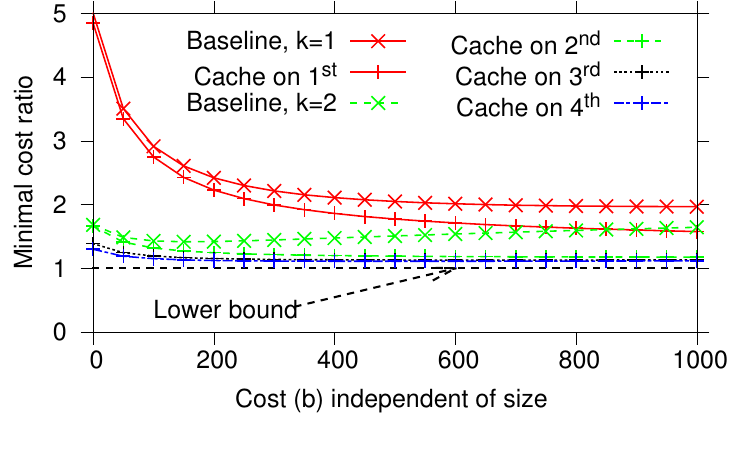}}
         \subfigure[Varying $H_{\min}$]{
           \includegraphics[trim = 1mm 10mm 4mm 0mm, width=0.30\textwidth]{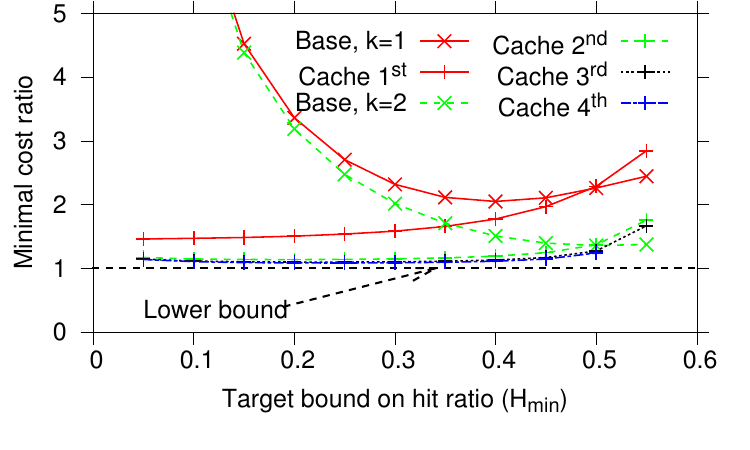}}
         \subfigure[Varying $\lambda_{\textrm{high}}$]{
           \includegraphics[trim = 1mm 10mm 4mm 0mm, width=0.30\textwidth]{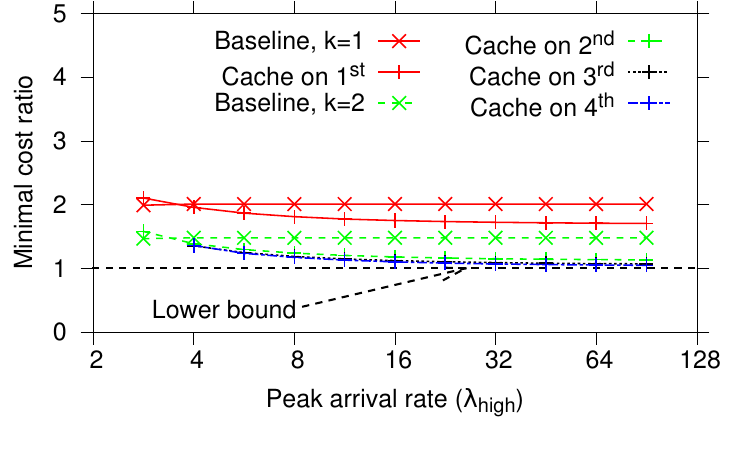}}
         \vspace{-8pt}
         \caption{Ratio of the minimal cost to the lower bound versus $b$, $H_{\min}$, and $\lambda_{\textrm{high}}$
           (other parameters at defaults).}
         \label{fig:dyn1}
         \end{figure*}
         \begin{figure*}[t]
         \centering
         \subfigure[Varying $H_{\min}$, $\alpha = 0.5$, $N = 10,000$]{
           \includegraphics[trim = 1mm 10mm 4mm 0mm, width=0.30\textwidth]{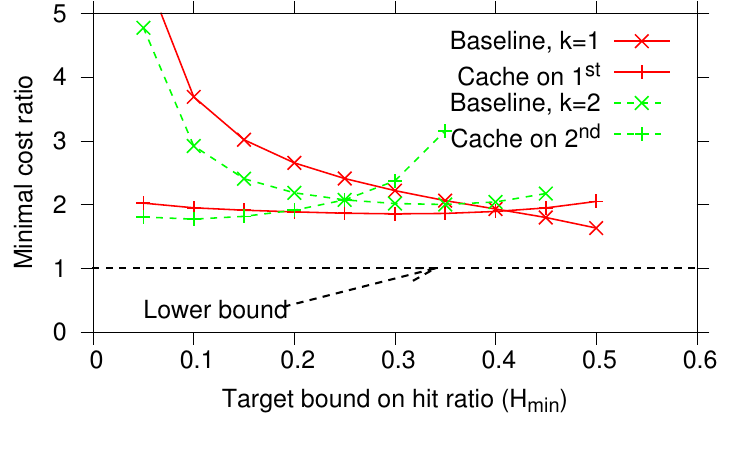}}
         \subfigure[Varying $H_{\min}$, $\alpha = 1$, $N = 10,000$]{
           \includegraphics[trim = 1mm 10mm 4mm 0mm, width=0.30\textwidth]{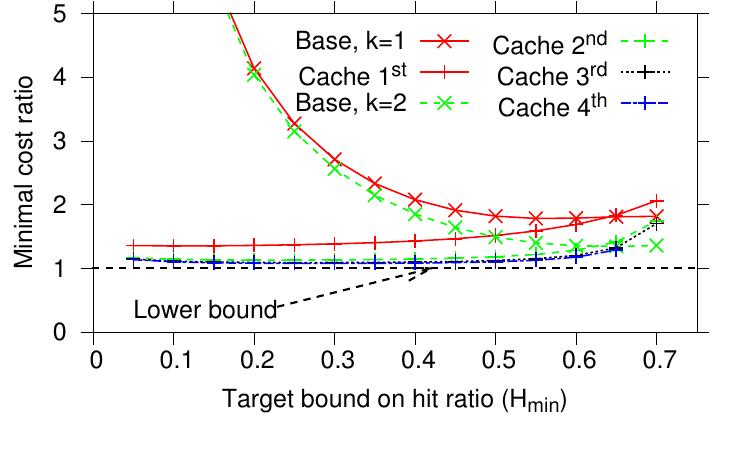}}
         \subfigure[Varying $h$; $\lambda_{\textrm{low}}$ = 0.1, $\lambda_{\textrm{high}}$, $\lambda_{\textrm{high}}$
           scaled so req. vol. unchanged]{
           \includegraphics[trim = 1mm 10mm 4mm 0mm, width=0.30\textwidth]{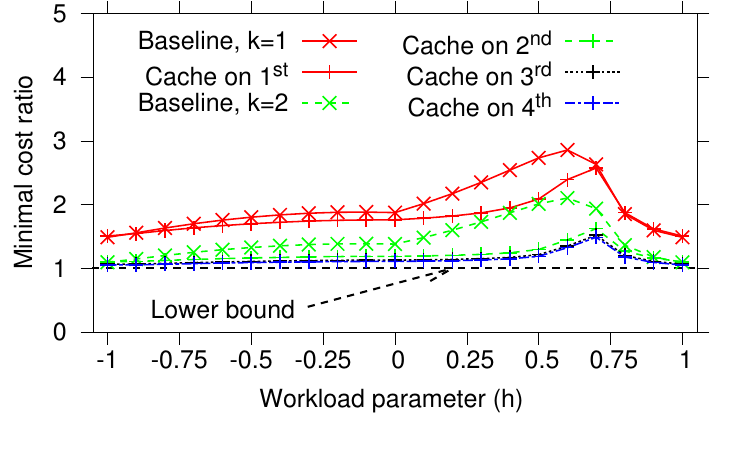}}
         \vspace{-11pt}
         \caption{Impact of popularity skew, number of objects, and request rate variability.}
         \label{fig:dyn2}
       \end{figure*}
       
       \begin{figure*}[t]
         \centering
         \subfigure[Varying $b$, Baseline $k=1$]{
           \includegraphics[trim = 1mm 10mm 4mm 0mm, width=0.23\textwidth]{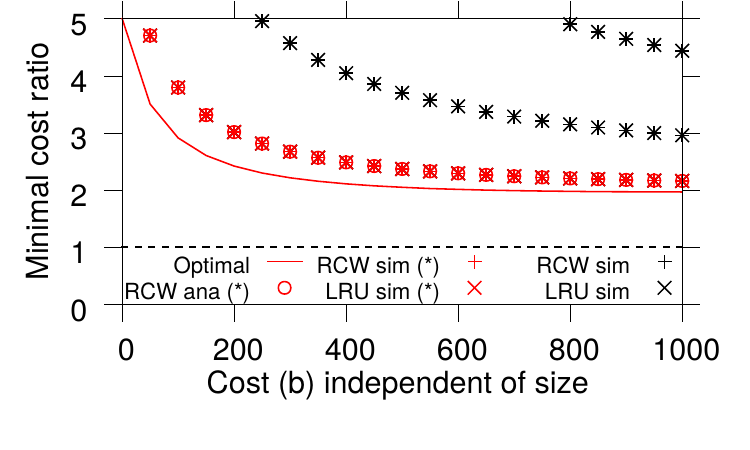}}
           \subfigure[Varying $b$, Cache on 1$^{st}$]{
           \includegraphics[trim = 1mm 10mm 4mm 0mm, width=0.23\textwidth]{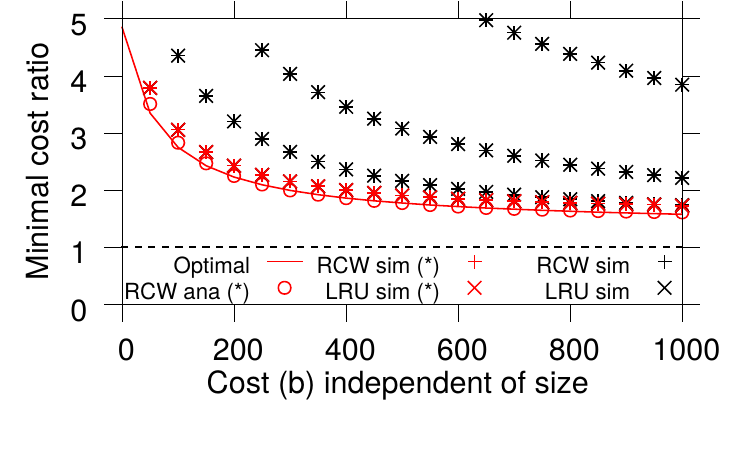}}
           \subfigure[Varying $b$, Cache on 2$^{nd}$]{
           \includegraphics[trim = 1mm 10mm 4mm 0mm, width=0.23\textwidth]{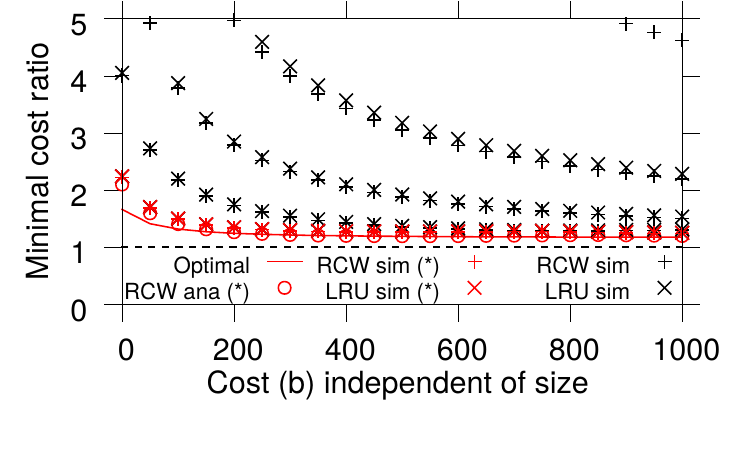}}
           \subfigure[Varying $b$, Cache on 4$^{th}$]{
           \includegraphics[trim = 1mm 10mm 4mm 0mm, width=0.23\textwidth]{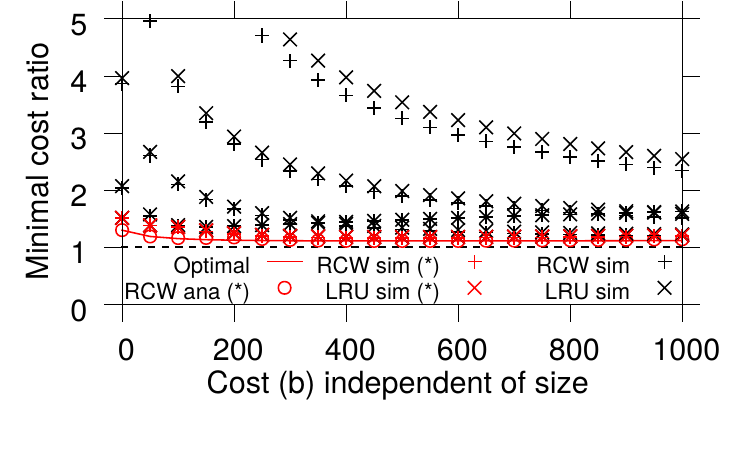}}
         \subfigure[Varying $H_{\min}$, Baseline $k=1$]{
           \includegraphics[trim = 1mm 10mm 4mm 0mm, width=0.23\textwidth]{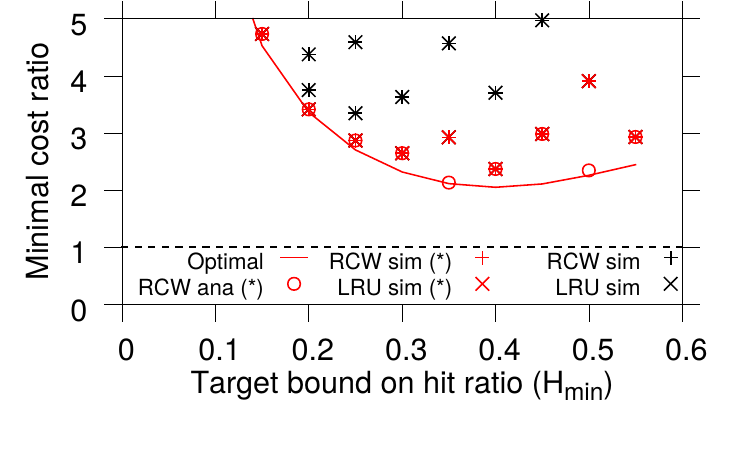}}
           \subfigure[Varying $H_{\min}$, Cache on 1$^{st}$]{
           \includegraphics[trim = 1mm 10mm 4mm 0mm, width=0.23\textwidth]{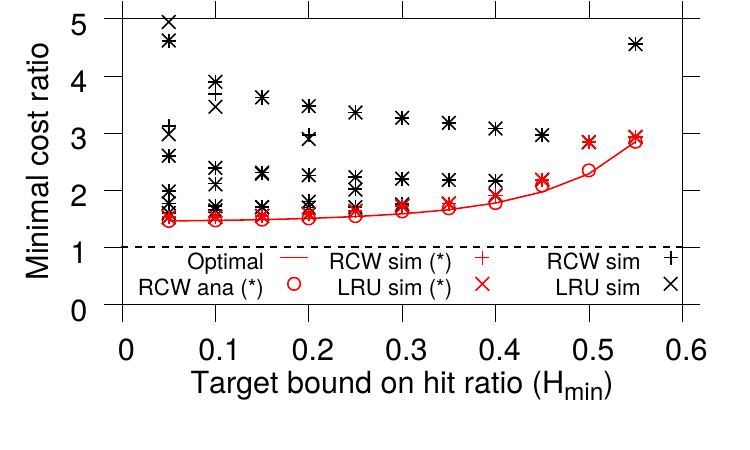}}
           \subfigure[Varying $H_{\min}$, Cache on 2$^{nd}$]{
           \includegraphics[trim = 1mm 10mm 4mm 0mm, width=0.23\textwidth]{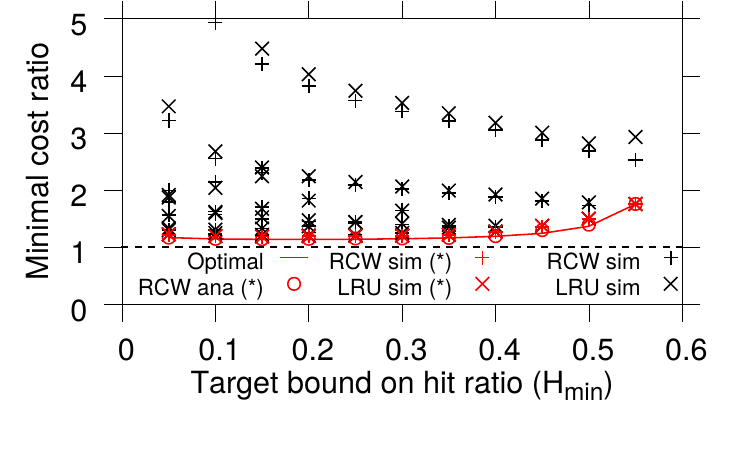}}
           \subfigure[Varying $H_{\min}$, Cache on 4$^{th}$]{
           \includegraphics[trim = 1mm 10mm 4mm 0mm, width=0.23\textwidth]{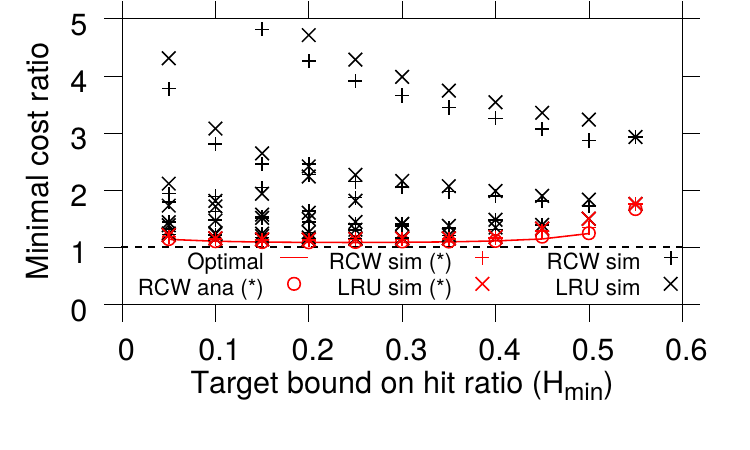}}
           \subfigure[Varying $\lambda_{\textrm{high}}$, Baseline $k=1$]{
           \includegraphics[trim = 1mm 10mm 4mm 0mm, width=0.23\textwidth]{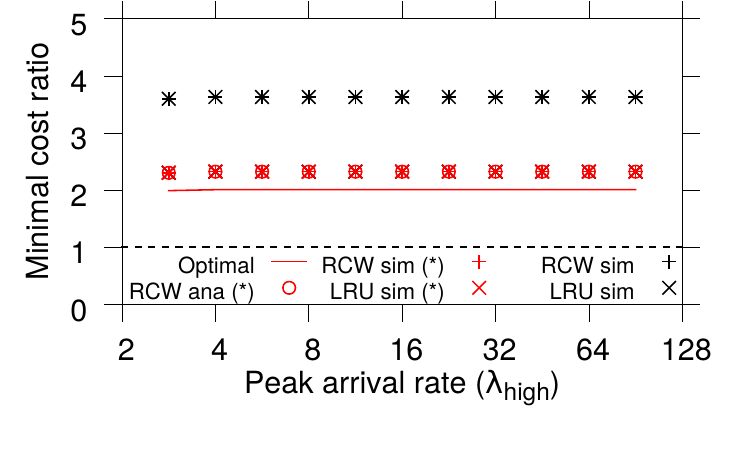}}
           \subfigure[Varying $\lambda_{\textrm{high}}$, Cache on 1$^{st}$]{
           \includegraphics[trim = 1mm 10mm 4mm 0mm, width=0.23\textwidth]{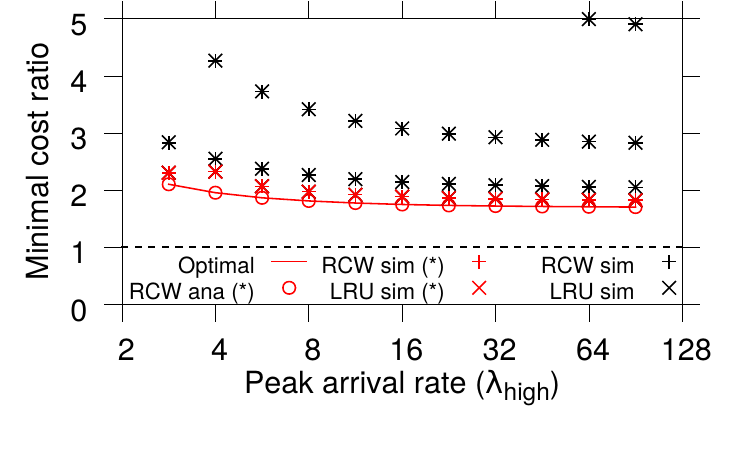}}
           \subfigure[Varying $\lambda_{\textrm{high}}$, Cache on 2$^{nd}$]{
           \includegraphics[trim = 1mm 10mm 4mm 0mm, width=0.23\textwidth]{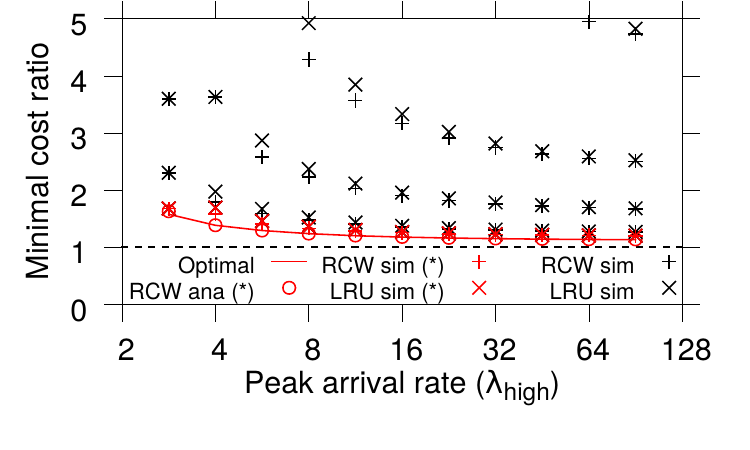}}
           \subfigure[Varying $\lambda_{\textrm{high}}$, Cache on 4$^{th}$]{
           \includegraphics[trim = 1mm 10mm 4mm 0mm, width=0.23\textwidth]{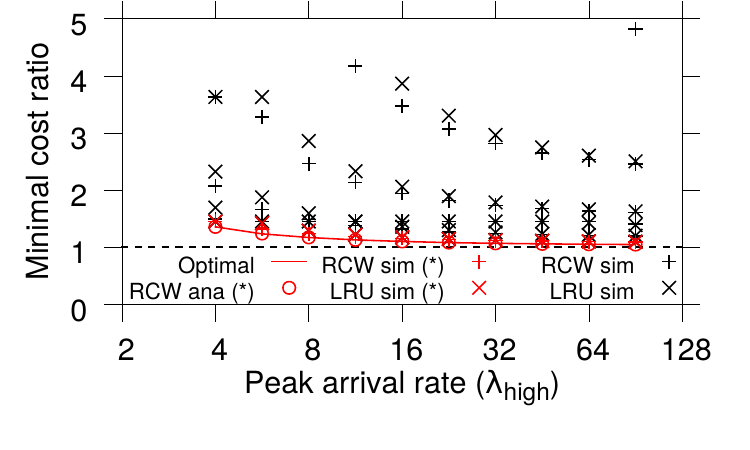}}
         \vspace{-8pt}
         \caption{\revfour{}{Demonstrating the effects of optimizing using only sample simulations (or analytic evaluations) based on a smaller set of cache size $C$ rather than optimizing over all possible $C$, $t_a$, $t_d$. Here, we show the optimized values for the example configurations that achieve a ratio of the minimal cost to the lower bound that is less than five, when using one of the following (discrete) cache sizes $C=10, 20, 46, 100, 200, 460, 1000, 2000, 4600, 10000$. 
         For each x-value of interest, results for the best of these configurations are shown using red markers, while results for the others are shown with black markers.  Results from both RCW simulations ($+$) and LRU simulations ($\times$) are shown.
         We also include the best (over these $C$ values) locally optimized delivery cost ratio when using the the RCW approximation ($\circ$) and the global optimal value from Figure 4 (red line) for each policy, which is optimized across all possible $C$, $t_a$, $t_d$.  The rows in the sub-figure matrix show the \revDerek{}{ratio of the} 
         minimal cost to the lower bound versus $b$, $H_{\min}$, and $\lambda_{\textrm{high}}$ (other parameters at defaults).  The columns show the results for four example policies.}}
         \label{fig:sample-based}
         \end{figure*}

\section{Dynamic Instantiation Performance}\label{comparisons}

%
%
For an initial model of request rate variation, we use a
single-parameter model in which the
request rate increases
linearly from a rate of zero at the beginning of the
time period to a rate $\lambda_{\textrm{high}}$
half-way through, and
then decreases linearly such that the request rate at the end of
the period is back to zero.
Default parameter settings (each used unless otherwise stated) are
$T$ = 1440 min. (24 hours),
$\lambda_{\textrm{high}} = 20$ req./min., $b = 500$ (and so for a cache
capacity of 1000 objects, for example, the size-independent portion of the
cache cost contributes half of the total), $H_{min} = 0.4$,
$N = 100,000$,
and a Zipf object popularity
distribution with $\alpha = 1$.

Figures~\ref{fig:dyn1}(a), (b), and (c) show the ratio of the minimal cost
for a dynamically instantiated cache
using different cache insertion policies (using $W$$=$$L$ for the
\emph{Cache on $k^{th}$ request} policies) to the cost lower bound,
as obtained from numerically solving the
optimization models of Section~\ref{optimizationmodel}, as a function of 
the cost parameter $b$, the hit rate constraint $H_{\min}$, and
the peak request rate $\lambda_{\textrm{high}}$, respectively.
\revfour{Also shown are}{As {\em baseline} comparisons, we also include}
the cost ratios for \emph{Cache on $1^{st}$ request}
\revfour{}{($k=1$)}
and \emph{Cache on $2^{nd}$ request} 
\revfour{}{($k=2$)}
for the baseline case of a
permanently allocated 
\revfour{cache with hit rate $H_{\min}$.}{LRU cache with capacity $C$ chosen so as to yield a hit rate of $H_{min}$ (as calculated from our RCW analysis).}
\revrev{In each figure}{In each figure,}
all other parameters are set to their default values.

Note that in these results:
(1) unless $b$ is very small (in which case, it is most cost-effective
to permanently allocate a small cache), $H_{\min}$ is large,
or $\lambda_{\textrm{high}}$ is too small for a dynamically instantiated
cache to fill,
dynamic cache instantiation can yield substantial cost savings;
(2) \emph{Cache on $k^{th}$ request} for $k \geq 2$
provides a better cost/performance tradeoff curve
compared to \emph{Cache on $1^{st}$ request};
and (3) there is only modest
room for
\revthree{improvement in cost/performance}{further improvement}
through use of
more complex cache insertion and replacement policies.

\revfour{}{The results also provide some interesting observations with regards to a permanently allocated LRU cache (with optimized cache capacity $C$).  For example, in Figure~\ref{fig:dyn1}(b) the curve for the {\em baseline with $k = 1$} goes down first, since for very low values of $H_{min}$ it is very inefficient to permanently allocate a cache.  As $H_{min}$ increases from these very low values, permanently allocating a cache becomes more reasonable.  Eventually, however, the baseline curve starts to rise.  In this region, permanently allocating a cache is reasonable.  However, the required cache capacity $C$ for the {\em baseline with $k=1$} begins to dominate the parameter $b$ (which gives the portion of the cost that is independent of cache size) in the cost expression, and so the inefficiency of a {\em Cache on 1$^{st}$ request} LRU cache in comparison to a cache in which the most popular objects are always kept in the cache becomes more and more apparent (as $C$ gets bigger and bigger relative to $b$), causing the baseline curve to rise.}

The potential benefits of dynamic cache instantiation
\revthree{(as well as of caching itself)}{(and of caching itself)}
are strongly dependent on the popularity skew.  
When object popularities follow a Zipf distribution with $\alpha$$=$$0.5$,
with our default
\revthree{parameter settings}{parameters}
it is not even possible to achieve
the target fraction of requests $H_{\min}$ to be served locally, using
dynamic cache instantiation.
This is partly due to the fact that
for $\alpha$$=$$0.5$, caching performance is degraded much more severely
when in the transient period than for $\alpha$$=$$1$
\revthree{(as seen by the analysis results in Section~\ref{transientperiodanalysis}),}{(e.g., results in Section~\ref{transientperiodanalysis}),}
and partly due to the fact that a larger cache is required to achieve
a given hit rate.
The impact of the popularity skew can be clearly seen by comparing the
results in Figures~\ref{fig:dyn2}(a) and (b), which use $N$$=$$10,000$
instead of the default value of 100,000 so as to allow the hit rate
constraint to be met over a significant range of values, even when
$\alpha$$=$$0.5$.
In addition to the poorer performance of dynamic cache
instantiation that is seen in Figure~\ref{fig:dyn2}(a),
note also the increased gap with respect to the lower bound,
and the poorer performance of \emph{Cache on $2^{nd}$ request} relative to
\emph{Cache on $1^{st}$ request} (compared to the relative performance
seen in Figure~\ref{fig:dyn2}(b)).
(Results for \emph{Cache on $k^{th}$ request} for $k = 3,4$ are not
shown in Figure~\ref{fig:dyn2}(a),
since the required value of $L$ becomes too large
for all but the smallest cache sizes.)

The significant impact of $N$ can be seen by comparing
Figures~\ref{fig:dyn1}(b) and~\ref{fig:dyn2}(b), which both use
$\alpha$$=$$1$ and differ only in the value of $N$.
\revfour{}{Here, we note that smaller $N$ (Figure~\ref{fig:dyn2}(b)) results in a smaller cache size $C$ being needed to achieve a given hit rate $H_{min}$.  This decreases the relative importance of $C$ compared to that of $b$ in the cost expressions, and correspondingly compresses the gaps between the curves with dynamic cache instantiation but different cache insertion policies.}

\revfour{Finally, the}{The} 
extent of rate variability also has a substantial impact.
This is illustrated in Figure~\ref{fig:dyn2}(c), for which
our model of request rate variation is modified so that
the minimum rate is $\lambda_{\textrm{low}}$,  $0$$<$$\lambda_{\textrm{low}}$$<$$\lambda_{\textrm{high}}$,
rather than zero, and with
linear rate increase/decrease occupying only a fraction $1$$-$$|h|$
of the time period, where $h$ is a parameter between $-$1 and 1.
When $h$$>$$0$, the request rate is
$\lambda_{\textrm{low}}$ for the rest of the time
period, while when $h$$<$$0$, the request rate is
$\lambda_{\textrm{high}}$ for the rest of the time
period (and so during the fraction $1$$-$$|h|$ of the time period
the rate first decreases linearly to $\lambda_{\textrm{low}}$ and then
increases linearly back to $\lambda_{\textrm{high}}$),
giving a peak to mean request rate ratio for $-1$$<$$h$$<$$1$ of
$2 / (1 - h + (1 + h)\lambda_{\textrm{low}}/\lambda_{\textrm{high}})$.
Results are shown for varying $h$, with $\lambda_{\textrm{low}}$ fixed
at 10\% of $\lambda_{\textrm{high}}$, and
$\lambda_{\textrm{high}}$ scaled for each value of $h$ so as to maintain
the same total request volume as with the default single-parameter model.
Note that $h$$=$$1$ and $h$$=$$-1$ correspond to the same scenario,
since in both cases the request rate is constant throughout the period.
Note also that the lower bound becomes overly optimistic for $h$ around 0.7;
in this case the requests are highly concentrated, and the
solution to the lower bound optimization problem is a large cache allocated
for a short period of time (for which the upper bound on hit rate when the
$\lfloor{C}\rfloor$ most popular objects are present in the cache becomes
quite loose).
Most importantly, observe that
when the pattern of request rate variation 
is such that there is a substantial ``valley"
\revrev{}{($h$$>$$0$)}
within the time period during
which the request rate is relatively low,
the benefits of dynamic instantiation
are much higher than when
there is a substantial ``plateau"
\revrev{within the time period
during which the request rate is relatively high.}{($h$$<$$0$).}

\revfour{}{
Finally, we look closer at the importance of optimizing both the cache size $C$ and the time period that the cache is dynamically allocated ($t_a$ to $t_d$).  For these experiments we compare the optimized results presented in Figure~\ref{fig:dyn1} with results obtained if using a smaller subset of (average) cache sizes $C=10, 20, 46, 100, 200, 460, 1000, 2000, 4600, 10000$.  Here, we include both simulation-based results and results obtained using our RCW approximations. In particular, Figure~\ref{fig:sample-based} shows the optimal values for the above $C$ configurations that are able to achieve an optimized cost that is less than five \revDerek{}{times} greater than the lower bound. Here, the rows in the sub-figure matrix show the \revDerek{}{ratio of the} minimal cost to the lower bound versus $b$, $H_{\min}$, and $\lambda_{\textrm{high}}$, respectively, and the columns show the results for four example policies ({\em Baseline $k=1$}, {\em Cache on 1$^{st}$ request}, {\em Cache on 2$^{nd}$ request}, and {\em Cache on 4$^{th}$ request}).
The figure clearly shows that the choice of $C$ is important.  For example, for each example policy (left-to-right) typically only a small subset of the $C$ choices achieved close to optimal cost for that policy (red line). However, even with only this small subset of possible caches sizes (separated by roughly a factor 2) we are able to achieve close to the best performance with each policy. This is illustrated by the red markers associated with RCW simulated ($+$), LRU simulated ($\times$), and RCW approximation ($\circ$) typically being close to the optimal policy version (red line).}

\revfour{}{We also note that the more cost-effective policies (that achieve closer to the lower bound) are less sensitive to the choice of $C$ compared to {\em Baseline, $k=1$} (that never \revDerek{turns off}{deallocates} the cache).  The instances of this policy with the biggest differences between the most cost-effective observed configuration (red markers) and the optimum configuration are primarily associated with instances where the best selected configuration (from the smaller subset of possible $C$ values) results in a significantly higher hit rate than $H_{min}$.  Of course, for these instances, a reduced cost (down to the optimal for that policy) could have been achieved by reducing the cache  size down to the smallest cache size that achieves $H_{min}$.  Even though our model is highly accurate for this policy (Sections 3.4 and 4.4) we also observe some instances where the observed differences between the best points (red markers) \revDerek{}{from analysis and simulations} are substantial. These instances corresponds to cases where for the same $C$, \revDerek{one}{} the analytic value ended up slightly above $H_{min}$ and the simulation values ended up slightly below $H_{min}$ (and therefore \revDerek{do not satisfy}{not satisfying} the optimization constraints), resulting in a different cache size $C$ being used for the \revDerek{simulations}{simulation points} than \revDerek{the analytic expression}{for the analysis}.  Of course, with more fine grained cache sizes the best values would be much more similar and closer to the optimal configuration (red line).}

\revfour{}{
For the simulation-based results presented here (similar to the analytic results), we used the simulation results from previous sections to obtain the cache hit rate for both transient and steady-state periods using each cache size $C$ of interest.  Then, we used knowledge of the workload to optimize $t_a$ and $t_d$ based on these values.  
Here, we simply find the largest $t_a$ 
\revDerek{$=T-t_d$ (by symmetry in the optimal solution of this workload)}{(with $t_d$ chosen as $T - t_a$, by symmetry in the optimal solution for this workload)} for which the expected hit rate is at least $H_{min}$.
This approach can of course be taken for any simulated/operational system for which one can measure both steady-state and transient hit rates and the number of requests that a transient period typically takes. We therefore argue that several of the optimization formulations that we use for optimized policies in this paper also could be applied when optimizing systems with harder-to-analyze policies as long as the system can be measured during both transient and steady-state for different cache sizes and the overall workload volume can be properly predicted over time.}

\section{Related Work}\label{relatedwork}

Caching policies typically either evict objects from the cache when the cache becomes full
(i.e., capacity-driven policies~\cite{PoBo03,BaOb00,DaTo90,ACD+00,JuBB03,BaMa05})
or based on the time since each individual object was last accessed
or entered the cache (i.e., timing-based policies~\cite{JuBB03,BaMa05}).
In practice, use of capacity-based policies such as LRU have dominated.
Unfortunatly, these policies are extremely hard to analyze exactly (e.g.,~\cite{King71, Gele73}).
This prompted the development of approximations~\cite{DaTo90,BCF+99,RoKT10}
and analysis of asymptotic properties~\cite{Fill96,Jele99,JeRa03,JeRa08,GKM+14}.
Most recent work use that the performance of capacity-driven policies often can be well
approximated using TTL-based caches~\cite{ChTW02, FrRR12, BDC+13, BGSC14, GaLM16, CaEa18}.
TTL-based models have also been used to analyze networks of caches~\cite{FNNT12, FDT+14, FNNT14, BGSC14, GaLM16},
to derive asymptotically optimized solutions~\cite{FeRP16},
for optimized server selection~\cite{CEGL14},
for utility maximization~\cite{DMT+16}, and for on-demand contract design~\cite{MaTo15}.

Few papers (regardless of replacement policy) have modeled discriminatory caching policies such as \emph{Cache on $k^{th}$ request}.
In our context, these policies are motivated
by the risk of cache pollution in small dynamically instantiated caches,
and more generally by the long tail of one-timers (one-hit wonders) observed
in edge networks~\cite{GALM07,ZSGK09, MaSi15,CaEa17}.
Recent works include trace-based evaluations
of \emph{Cache on $k^{th}$ request} policies~\cite{MaSi15,CaEa17},
simple analytic models for hit and insertion probabilities that (in contrast to us)
ignore cache replacement~\cite{CaEa17},
or works that have used TTL-based recurrence expressions to
model variations of \emph{Cache on $k^{th}$ request}~\cite{GaLM16, MaGL14,GaVa15,GaVa16,CaEa18}.
Of these works,
only Carlsson and Eager try to minimize the delivery cost~\cite{CaEa18}.
However, in contrast to us, they assume elastic TTL caches and only consider a single file.

Other related works have adapted the individual TTL values of each object so to achieve
some objective~\cite{GKKP19,BSG+17}.  For example,
Carra et al.~\cite{CaNM19} demonstrate how the individual TTL values
of each object can be adapted (with constant overhead)
so to closely track the optimal cache configurations.
However, these works only consider {\em Cache on 1$^{st}$ request} policies.

To simplify analysis, TTL-based approximations of LRU-based caches
typically leverages the general ideas of a cache characterization time~\cite{ChTW02, FrRR12},
which in the simplest case corresponds to the (approximate) time that the object stays in the cache if not requested again.
This time corresponds closely to our parameter $L$,
with the important difference that the RCW caches use a {\em{request count window}} rather than a {\em{time window}}.
This subtle difference makes our approach favorable when modeling fixed capacity caches
in systems with substantial request rate variations (e.g., according to a diurnal cycle).
Furthermore, our RCW approach allows us to
derive (i) exact expressions
for general {\em Cache on k$^{th}$ request} policies,
popularity distributions, and transient periods,
and (ii) corresponding $\mathcal{O}(1)$ computational cost approximations.
Such
results, which we need for our optimization models,
are not found in the TTL-based modeling literature.

For the case of an infinite \emph{Cache on $1^{st}$ request} cache
with a finite request stream, Breslau et al.~\cite{BCF+99} provide
exact hit rate expression for general popularity distributions,
as well as approximate scaling properties for Zipf distributions.
However, we did not find these scaling relationships (focusing on orders)
sufficient for our analysis and therefore developed more precise expressions
for Zipf with $\alpha = 1$ and $\alpha = 0.5$.

The (general) idea of dynamically adapting the amount of dedicated resource based
on time-varying wokloads is not
\revthree{new.  In addition to a large number of auto-scaling techniques~\cite{LoML14,PAA+16}
for dynamically scaling ome elastic resource (e.g., front end servers) to match the overall request load
(e.g., by turning on/off servers or VMs in datacenters~\cite{LWAT13} or distributed server clusters~\cite{MaSS12}),
various approaches have been designed to combine and manage a collection of instances so to minimize the storage costs (e.g.,~\cite{XSDW16}).
These works typically assume a uniform workload and do not consider caching of the individual objects being requested.}{new~\cite{LWAT13,MaSS12,LoML14,PAA+16}.}
Within the context of cloud-based caching,
some recent works consider how to scale resources based on time varying workloads
and diurnal patterns~\cite{CaNM19,SuKS16,DaCa14}.
For example, in addition to the work by Carra et al.~\cite{CaNM19} (discussed above),
Sundarrajan et al.~\cite{SuKS16} use discriminatory caching algorithms together with
partitioned caches to save energy during off-peak hours, Dan and Carlsson~\cite{DaCa14}
optimize what objects to push to cloud storage based on diurnal demands
and a basic cost model, and Cai et al.~\cite{CaLK14} considered the problem of how to scale the number of
cache servers in a hierarchy of LRU caches.
Others have considered how CDNs best collaborate with ISPs making available microdatacenters~\cite{FPL+13},
how to build virtual CDNs on-the-fly on top of shared third-party infrastructures~\cite{KIM+17},
or have optimized the caching hierarchy of cloud caches
based on the assumption that request rates are
known and stationary (ignoring time-varying request loads)~\cite{MLTP16}.
None of these works consider the problem of
optimized
cache instantiation.

\nocite{kim2019novel,tan2020puf}

\section{Conclusions}\label{conclusions}

In this paper we have taken a first look at dynamic cache instantiation.
For this purpose, we have derived new analysis results for what we
term ``request count window'' (RCW)
\revthree{caches.
  These results,}{caches,}
including explicit,
exact expressions for cache performance metrics for
\emph{Cache on $k^{th}$ request}
RCW caches for general $k$, and new 
$\mathcal{O}(1)$
computational cost approximations for cache performance metrics
for Zipf popularity distributions with $\alpha$$=$$1$ and
\revthree{$\alpha = 0.5$, may be of interest in their own right.}{$\alpha$$=$$0.5$.
  These results are of interest in their own right,
  especially as the performance of RCW caches is shown to closely
  match the performance of the corresponding \emph{Cache on $k^{th}$ request} LRU caches.}
  \revfour{}{This contribution is particularly important since it is the first to accurately capture the performance of LRU caches under time-varying workloads.}
We then applied our analysis results to develop optimization models
for dynamic cache instantiation parameters, specifically the
cache size and the duration of the cache instantiation interval,
for different cache insertion policies, as well as for
a cost lower bound that holds for {\em any} caching policy.
Our results
\revthree{suggest that
unless the component of cache cost that is independent of size
is very small, or it is necessary to achieve a high hit rate target
that is difficult or impossible to achieve with a dynamically instantiated
cache, or there is insufficient peak load or variability in load,}{show that}
dynamic cache instantiation using a selective cache insertion policy such
as \emph{Cache on $2^{nd}$ request} may yield substantial benefits
compared to a permanently allocated
\revthree{cache, and may be}{cache, and is}
a promising approach for content delivery applications.
\revthree{\revrev{Our}{Finally, we note that our}
work has used the independent reference model.
\revthree{Features commonly found in real workloads such as short-term temporal
locality, non-stationary object popularities, and high rates of addition
of new content should make dynamic cache instantiation potentially more
appealing, since they make yesterday's cache contents less useful.}{However,
  features commonly found in real workloads
  (e.g., short-term temporal locality, non-stationary object popularities, and high rates of addition of new content)
  make yesterday's cache contents less useful. The reported improvements are therefore conservative.
  In practice, dynamic cache instantiation is expected to be even more appealing.}}{Finally, we note that the use
  of the independent reference model (used here) provides conservative estimates of the potential
  improvements using dynamic instantiation, since in practice short-term temporal locality, non-stationary object popularities,
  and high rates of addition of new content typically would make yesterday's cache contents less useful.}
\revthree{\revthree{A promising direction for future work is to explore the potential
of dynamic cache instantiation using models that incorporate such features,
and/or trace-driven simulation using traces from real systems.}{Although we leave
  analysis of such workloads for future work (due to page limitations),
  this claim has been validated using trace-based simulations using a 20-month trace
  of YouTube requests on a campus network.}}{}

\section*{Acknowledgements}

This work was supported by funding from
the Swedish Research Council (VR)
and the Natural Sciences and Engineering Research Council (NSERC) of Canada.



{
  \bibliographystyle{IEEEtran}
  \bibliography{paper-journal}
}

%

\begin{IEEEbiography}[{\includegraphics[width=1in,height=1.55in,clip,keepaspectratio]{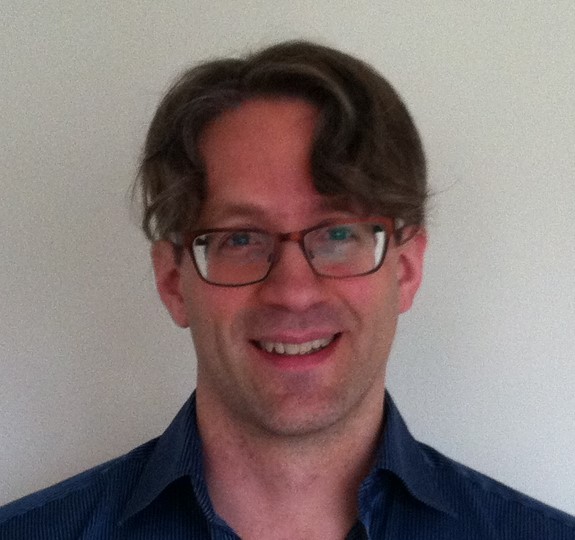}}]{Niklas Carlsson}
  is an Associate Professor at Link\"oping
  University, Sweden. He received his M.Sc. degree in Engineering
  Physics from Ume\aa~University, Sweden, and his Ph.D. in Computer
  Science from the University of Saskatchewan, Canada. He has
  previously worked as a Postdoctoral Fellow at the University
  of Saskatchewan, Canada, and as a Research Associate at the
  University of Calgary, Canada. His research interests are in the
  areas of design, modeling, characterization, and performance
  evaluation of distributed systems and networks.
\end{IEEEbiography}

\begin{IEEEbiography}[{\includegraphics[trim=100mm 0mm 140mm 0mm, clip, width=1in, clip, keepaspectratio]{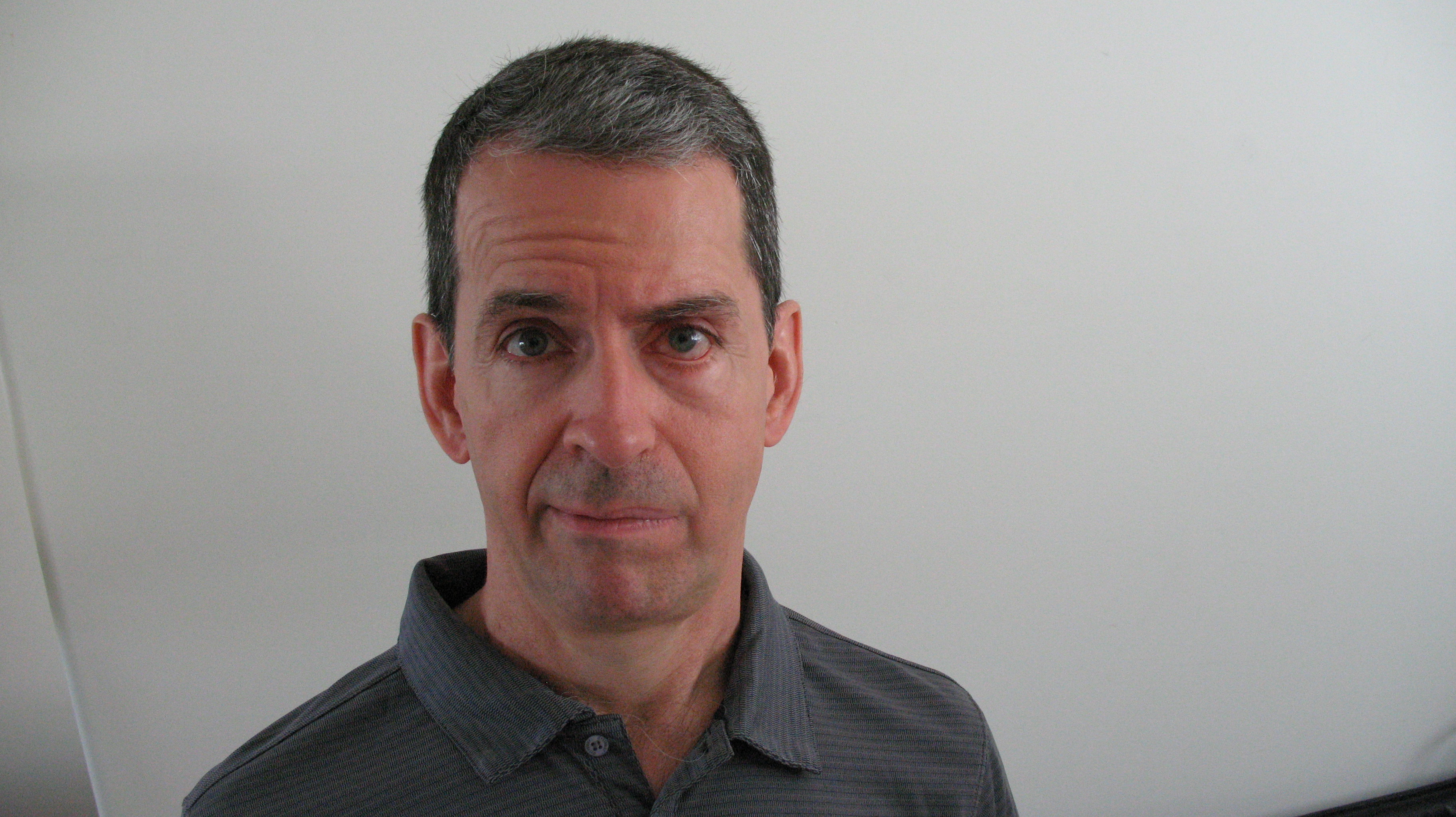}}]{Derek Eager}
  received the BSc degree in computer science from the University of Regina, Canada,
  and the MSc and PhD degrees in computer science from the University of Toronto, Canada.
  He is a professor in the Department of Computer Science, University of Saskatchewan, Canada.
  His research interests include the areas of performance evaluation, content distribution,
  and distributed systems and networks.
\end{IEEEbiography}
  







\end{document}